\documentclass[10pt,final,twoside]{IEEEtran}

\RequirePackage[table,svgnames]{xcolor} 
\RequirePackage{ifxetex}
\usepackage[normalem]{ulem} 
\usepackage{calc}

\usepackage{tikz} 
\usetikzlibrary{svg.path}
\usepackage{scalerel}
\definecolor{orcidlogocol}{HTML}{A6CE39}
\tikzset{
  orcidlogo/.pic={
    \fill[orcidlogocol] svg{M256,128c0,70.7-57.3,128-128,128C57.3,256,0,198.7,0,128C0,57.3,57.3,0,128,0C198.7,0,256,57.3,256,128z};
    \fill[white] svg{M86.3,186.2H70.9V79.1h15.4v48.4V186.2z}
                 svg{M108.9,79.1h41.6c39.6,0,57,28.3,57,53.6c0,27.5-21.5,53.6-56.8,53.6h-41.8V79.1z M124.3,172.4h24.5c34.9,0,42.9-26.5,42.9-39.7c0-21.5-13.7-39.7-43.7-39.7h-23.7V172.4z}
                 svg{M88.7,56.8c0,5.5-4.5,10.1-10.1,10.1c-5.6,0-10.1-4.6-10.1-10.1c0-5.6,4.5-10.1,10.1-10.1C84.2,46.7,88.7,51.3,88.7,56.8z};
  }
}   
\newcommand\orcidicon[1]{\,\,\href{https://orcid.org/#1}{\mbox{\scalerel*{
\begin{tikzpicture}[yscale=-1,transform shape]
\pic{orcidlogo};
\end{tikzpicture}
}{|}}}}

\AtBeginDocument{\hypersetup{pdftitle={Flexible Fast-Convolution Processing for Cellular Radio Evolution}}}
 
\usepackage{graphicx}   
\DeclareRobustCommand*{\IEEEauthorrefmark}[1]{\raisebox{0pt}[0pt][0pt]
{\textsuperscript{\footnotesize\ensuremath{\ifcase#1\or 1\or 2\or 3\or%
    4\or 5\or 6\or 7\or 8
\else\textsuperscript{\expandafter\romannumeral#1}\fi}}}}
\usepackage[table,svgnames]{xcolor} 

\vbadness=\maxdimen

\xdefinecolor{captCol}{cmyk}{0.86, 0.35, 0, 0.35}
\xdefinecolor{spotColor}{cmyk}{1.0, 0.28, 0.00, 0.21}
\xdefinecolor{sectColor}{cmyk}{0.0, 0.11, 0.09, 0.86}
\xdefinecolor{Chart1}{HTML}{004586}
\xdefinecolor{Chart2}{HTML}{FF420E}
\xdefinecolor{Chart3}{HTML}{FFD320} 
\xdefinecolor{Chart4}{HTML}{579D1C}  
\xdefinecolor{Chart5}{HTML}{7E0021}
\xdefinecolor{Chart6}{HTML}{83CAFF}
\xdefinecolor{Chart7}{HTML}{314004}
\xdefinecolor{Chart8}{HTML}{AECF00}
\xdefinecolor{Chart9}{HTML}{4B1F6F}
\xdefinecolor{Chart10}{HTML}{FF950E}
\xdefinecolor{Chart11}{HTML}{C5000B}
\xdefinecolor{Chart12}{HTML}{0084D1}

\usepackage{ifxetex} 
\ifxetex  
  \usepackage{fourier-otf}
  \usepackage{fontspec}
  \usepackage{xltxtra}

\else
  \usepackage{amsmath,amssymb}
  \usepackage{stix}
\fi  

\usepackage{mathtools}
\usepackage{mathdots}
\DeclarePairedDelimiter{\diagpars}{(}{)}

\DeclarePairedDelimiter{\vecpars}{(}{)}
\DeclarePairedDelimiter{\norm}{\lVert}{\rVert}

\newcommand{\diag}{\operatorname{diag}\diagpars}
\newcommand{\bdiag}{\operatorname{bdiag}\diagpars}

\renewcommand{\vec}{\operatorname{vec}\vecpars}

\usepackage[   
  draft=false,
  colorlinks=true,  
  breaklinks=true,
  allcolors=DarkBlue,
  urlbordercolor={1 0 0}, 
  pdfborderstyle={/S/U/W 0.5},
  ]{hyperref}

\hypersetup{%
  pdfsubject={This paper proposes new symbol-synchronized fast-convolution-based scheme for flexible and effective waveform generation and processing in 5th generation (5G) systems.},
  pdfkeywords={5G, physical layer, 5G New Radio, 5G-NR, multicarrier, waveforms, filtered-OFDM, fast-convolution},
  pdfauthor={Juha Yli-Kaakinen, Toni Levanen, Arto Palin, Markku Renfors, and Mikko Valkama}%
}
\hypersetup{pdflinkmargin=0.5pt,hidelinks}

\usepackage[version-1-compatibility]{siunitx}
\DeclareSIUnit\prb{\text{PRB}}
\DeclareSIUnit\prbs{\text{PRBs}}
 
\usepackage{booktabs} 
\usepackage{cellspace}
\usepackage{cite}
  
\definecolor{pblue}{HTML}{0072BD}
\definecolor{porange}{HTML}{D95319} 
\definecolor{pyellow}{HTML}{EDB120}
\definecolor{ppurple}{HTML}{7E2F8E}
\definecolor{pgreen}{HTML}{77AC30}
\definecolor{pred}{HTML}{A2142F}

\providecommand{\iu}{\ensuremath{{\mathop{\mspace{1mu}\mathrm{j}\mspace{0.5mu}}\nolimits}}}

\usepackage[framemethod=TikZ]{mdframed}

\makeatletter
\newcommand*{\transpose}{%
  {\mathpalette\@transpose{}}%
}
\newcommand*{\htranspose}{%
  {\mathpalette\@htranspose{}}%
}
\newcommand*{\@transpose}[2]{%
  \raisebox{\depth}{$\m@th#1\mathsf{T}$}%
}
\newcommand*{\@htranspose}[2]{%
  \raisebox{\depth}{$\m@th#1\mathsf{H}$}%
}

\makeatother
\usepackage[nolist,nohyperlinks]{acronym}  

\makeatletter
\DeclareMathSizes{\@xpt}{\@xpt}{7}{5}
\makeatother
 
\makeatletter
\newif\ifonecolumn
\@ifclasswith{IEEEtran}{onecolumn}{\onecolumntrue}{\onecolumnfalse}
\makeatother

\newlength{\figwidth}
\newlength{\figwidthSC}
\ifonecolumn
\else
\fi
\begin{document} 

\title{Flexible Fast-Convolution Processing for Cellular Radio Evolution}

\author{%
  Juha Yli-Kaakinen\orcidicon{0000-0002-4665-9332},
  Toni Levanen\orcidicon{0000-0002-9248-0835}, %
  Arto Palin\orcidicon{0000-0001-8567-5549}, \\%
  Markku Renfors\orcidicon{0000-0003-1548-6851}, and %
  Mikko Valkama\orcidicon{0000-0003-0361-0800}
  \thanks{
    This work was supported in part by the Business Finland (formerly known as the Finnish Funding Agency for Innovation, Tekes) and Nokia Bell Labs through the Projects 5G-VIIMA and 5G-FORCE; in part by Nokia Networks; and in part by the Academy of Finland under Project 284694, Project 284724, and Project 319994. %
    (Corresponding author: Juha Yli-Kaakinen.)}  \thanks{Juha Yli-Kaakinen, Markku Renfors, and Mikko Valkama are with the Department of Electrical Engineering, Tampere University, FI-33101 Tampere, Finland (e-mail: $\lbrace$juha.yli-kaakinen; markku.renfors; mikko.valkama$\rbrace$@tuni.fi)} \thanks{Toni Levanen and Arto Palin are with Nokia Networks, 33100 Tampere, Finland (e-mail: $\lbrace$toni.a.levanen; arto.palin$\rbrace$@nokia.com)} 
}

\maketitle
\vspace{-3em}
\begin{abstract}
  Orthogonal frequency-division multiplexing (OFDM)\acused{ofdm} has been selected as a baseline waveform for \ac{lte} and \ac{5gnr}. Fast-convolution (FC)-based\acused{fc} frequency-domain signal processing has been considered recently as an effective tool for spectrum enhancement of \ac{ofdm}-based waveforms. \ac{fc}-based filtering approximates linear convolution by effective \ac{fft}-based circular convolutions using partly overlapping processing blocks. In earlier work, we have shown that \ac{fc}-based filtering is a very flexible and efficient tool for filtered-\ac{ofdm} signal generation and receiver side subband filtering. In this paper, we present a symbol-synchronous \ac{fc}-processing scheme flexibly allowing filter re-configuration time resolution equal to one \ac{ofdm} symbol while  supporting tight carrier-wise filtering for \ac{5gnr} in mixed-numerology scenarios with adjustable subcarrier spacings, center frequencies, and subband bandwidths as well as providing co-exitence with \ac{lte}. The proposed scheme is demonstrated to support envisioned use cases of \ac{5gnr} and provide flexible starting point for sixth generation development.
\end{abstract}    
 
\acresetall

\section{Introduction}    
\label{sec:introduction}\bstctlcite{IEEEexample:BSTcontrol}
\IEEEPARstart{O}{rthogonal} frequency-division multiplexing (OFDM)\acused{ofdm} is utilized in \ac{lte} and \ac{5gnr}\acused{5g-nr} due to its high flexibility and efficiency in allocating spectral resources to different users, simple and robust way of channel equalization, as well as simplicity of combining multiantenna schemes with the core physical-layer processing \cite{B:Dahlman2018}. The poor spectral localization of the \ac{ofdm}, however, calls for enhancements such as windowing or filtering to improve the localization of the waveform by effectively suppressing the unwanted emissions. This is important especially in challenging new spectrum use scenarios like asynchronous multiple access, as well as in mixed-numerology cases aiming to use adjustable symbol  and \ac{cp} lengths, \acp{scs}, and frame structures depending on the service requirements \cite{J:20145GNOW,J:2014BanelliModFormatsAndWaveformsFor5G,J:Choi19:trans_desig,J:Mao:FOFDM,J:Memisoglu2020}. 
   
Fast-convolution (\acs{fc})-based\acused{fc} filtering has been recently proposed as an efficient tool for spectrum control of singe-carrier and multi-carrier waveforms \cite{J:Boucheret99,J:Borgerding06, C:Tanabe09:FFT_FB,C:Renfors2015:fc-f-ofdm,C:David16:perfor_fbmc_oqam,J:Yli-Kaakinen:JSAC2017, C:Yli-Kaakinen:ASILOMAR2018,J:Li19:FC,J:Loulou19:advan_low_compl_multic_schem,J:Yli-Kaakinen:TWC2021, C:ishibashi21,J:Lin2021:HybricCarr}.
In general, the objective of the filtering is to improve the spectral utilization of the channel by improving the localization of the waveform in frequency direction, that is, maximizing the transmission bandwidth for a given channel bandwidth. \ac{fc}-based filter-bank solutions have superior flexibility when compared with the conventional polyphase-type filter banks \cite{J:Renfors14:FC}. \ac{fc} processing approximates a linear (aperiodic) convolution through effective \ac{fft}-based circular convolutions using partly overlapping processing blocks (so-called \ac{fc} blocks).  With \ac{fc} processing, it is very straightforward to adjust the bandwidths and the center frequencies of the subbands with possibly different numerologies individually \cite{J:Yli-Kaakinen:JSAC2017} or even at the symbol level.
 
In original \emph{continuous} \ac{fc}-based filtered-\ac{ofdm} processing model derived in \cite{C:Renfors2015:fc-f-ofdm,J:Yli-Kaakinen:JSAC2017}, continuous stream of \ac{cp}-\ac{ofdm}\acused{cp-ofdm} symbols are divided into overlapping \ac{fc}-processing blocks of the same length and the overlap between \ac{fc} blocks is fixed (typically \SI{50}{\%}). Since the \ac{cp} length in \ac{5gnr} is non-zero (and both the \ac{ofdm} symbol length and the \ac{fc}-processing block length typically take power-of-two values), \ac{fc} blocks are not time synchronized to \ac{cp-ofdm} symbols. The drawback of this approach is that, when the filter configuration changes, i.e., bandwidth or center frequency of the subband (or \ac{bwp} in the \ac{5g-nr} terminology) is modified, or for some other reason filtering parameters needs to be adjusted between two \ac{ofdm} symbols, then this change typically occurs within a \ac{fc}-processing block degrading the performance of the filtering during this block. 

In \emph{discontinuous symbol-synchronized} \ac{fc} processing as detailed in \cite{C:Yli-Kaakinen:ASILOMAR2018,J:Yli-Kaakinen:TWC2021}, each symbol is divided into fixed number of processing blocks (e.g. two). These \ac{fc}-processing blocks are then filtered using \ac{fc}-based circular convolutions and the filtered \ac{fc}-processing blocks are concatenated by using \ac{ola} processing to form a stream of filtered \ac{cp-ofdm} symbols. In this case, the change in filtering configuration does not induce any additional intrinsic interference, since the \ac{ofdm} symbol boundaries are also boundaries of the payload part of the \ac{fc} blocks. However, for this approach, the \ac{fc} processing is aligned only with one numerology at the time which may cause problems in supporting mixed numerology. Also, the needed \ac{ola} scheme may introduce additional constrains in time-critical applications due to the overlapping needed at the output side.

In the \emph{continuous symbol-synchronized} processing model proposed in this article, the continuous stream of symbols is divided into overlapping blocks such that the overlap is dynamically adjusted based on the \ac{cp} lengths thus guaranteeing the synchronous processing of all \ac{cp-ofdm} symbols for all numerologies with normal \ac{cp}.  Therefore, the proposed approach avoids the drawbacks of the original continuous and discontinuous symbol-synchronized \ac{fc}-based filtered-\ac{ofdm} models. The only drawback is that the smallest possible forward transform length is somewhat higher when compared to earlier approaches.

The  main  contributions  of this manuscript can be itemized as follows: 
\begin{itemize}
\item[\tiny$\blacktriangleright$] Proposed processing is optimized to \ac{5gnr} and \ac{lte} physical-layer numerologies, where all supported subcarrier spacings align in time with \SI{0.5}{ms} time resolution.
\item[\tiny$\blacktriangleright$] \ac{fc} blocks are aligned with \ac{ofdm} symbols of all different subcarrier spacings in mixed-numerology implementation of \ac{5gnr}.
\item[\tiny$\blacktriangleright$] \ac{fc} blocks are also aligned between \ac{lte} and all numerologies with \ac{5gnr}, allowing smooth carrier combining processing in multi-technology or multi-radio \ac{tx} and corresponding carrier separating \ac{rx} processing.
\item[\tiny$\blacktriangleright$] Only one \ac{fc} block within the \SI{0.5}{ms} time window has different overhead, all other \ac{fc} blocks have common overhead.
\item[\tiny$\blacktriangleright$] Processing can be done using either \ac{ols} or \ac{ola}, or even mix of these, providing additional degree of flexibility for implementation.
\item[\tiny$\blacktriangleright$] By using \ac{fc}-processing bin spacing of \SI{60}{kHz}, we can support dynamic changes in the filter parameterization with time resolution corresponding to the \SI{60}{kHz} subcarrier spacing \ac{ofdm} symbols.
\end{itemize}

The presented solution supports all different use cases envisioned for the flexible, \ac{bwp}-based \ac{5g-nr} radio interface, allowing filter re-configuration time resolution equal to one \ac{ofdm} symbol while maintaining high quality separation of different frequency blocks. The solutions presented here are especially important for below-\SI{7}{GHz} communications due to scarce spectral resources, but there is no limitation in applying the solutions also for higher carrier frequencies if seen necessary. 

The remainder of this paper is organized as follows. Section~\ref{sec:5Gnumerology} shortly reviews the \ac{5gnr} numerology and relevant terminology for reference. Then, the proposed continuous symbol-synchronized \ac{fc}-based filtered-\ac{ofdm} processing models for \ac{tx} and \ac{rx} are described in Section~\ref{sec:fc-f-ofdm}. This section also describes how to define the frequency-domain windows for reducing the \acp{oobe} and \ac{ini}. Section~\ref{sec:wavef-requ} introduces the key metrics and requirements used for evaluating the performance of the \ac{tx} processing. In Section \ref{sec:examples}, the performance of the proposed processing is demonstrated in various mixed-numerology scenarios. Finally, the conclusions are drawn in Section \ref{sec:conclucions}. 

\makeatletter
\def\iddots{\mathinner{\mkern1mu\raise\p@
\vbox{\kern7\p@\hbox{.}}\mkern2mu
\raise4\p@\hbox{.}\mkern2mu\raise7\p@\hbox{.}\mkern1mu}}
\makeatother
\makeatletter
\def\Ddots{\mathinner{\mkern1mu\raise\p@
\vbox{\kern7\p@\hbox{.}}\mkern2mu
\raise4\p@\hbox{.}\mkern2mu\raise7\p@\hbox{.}\mkern1mu}}
\makeatother

\section*{Notation and Terminology}
In the following, boldface upper and lower-case letters denote matrices and column vectors, respectively. $\mathbf{0}_{q\times p}$ and $\mathbf{1}_{q\times p}$ are the $q\times p$ matrices of all zeros and all ones, respectively. $\mathbf{I}_{q}$ and $\mathbf{J}_{q}$ are the identity and reverse identity matrices of size $q$ as given by 
\begin{equation}
  \mathbf{I}_q =
  \begin{bmatrix}
  1 & 0 & \cdots & 0 \\
  0 & 1 & \cdots & 0 \\ 
  \vdots & \vdots & \ddots & \vdots \\
  0 & 0 & \cdots & 1  
  \end{bmatrix}
  \quad\text{and}\quad
  \mathbf{J}_q=
  \begin{bmatrix}
  0 & \cdots & 0 & 1 \\
  0 & \cdots & 1 & 0 \\  
  \vdots & \iddots & \vdots & \vdots \\
  1 & \cdots & 0 & 0  
  \end{bmatrix}\nonumber,
\end{equation}\noindent
respectively. The entry on the $i$th row and $j$th column of a $q\times p$ matrix $\mathbf{A}$ is denoted by $[\mathbf{A}]_{i,j}$ for $i\in\{0,1,\dots,q-1\}$ and $j\in\{0,1,\dots,p-1\}$ and $[\mathbf{A}]_j$ denotes the $j$th column of $\mathbf{A}$. For vectors, $[\mathbf{a}]_j$ denotes the $j$th element of $\mathbf{a}$. The column vector formed by stacking vertically the columns of $\mathbf{A}$ is $\mathbf{a}=\vec{\mathbf{A}}$. Furthermore, $\diag{\mathbf{a}}$ denotes a diagonal matrix with the  elements of $\mathbf{a}$ along the main diagonal. The transpose of matrix $\mathbf{A}$ or vector $\mathbf{a}$ is denoted by $\mathbf{A}^\transpose$ and $\mathbf{a}^\transpose$, respectively, while the corresponding conjugate transposes are denoted by  $\mathbf{A}^\mathsf{H}$ and $\mathbf{a}^\mathsf{H}$, respectively. The Euclidean norm is denoted by $\norm{\cdot}$ and $\lvert\cdot\rvert$ is the absolute value for scalars and cardinality for sets.
Element-wise (Hadamard) product is denoted by $\odot$. \ifonecolumn
\else The list of most common symbols is given in Table~\ref{tab:symbols}.
\fi

\ifonecolumn
\else
\begin{table}[t]
  \caption{List of Symbols. In these notations, subscript $m$ denotes the subband index and $n$ denotes the \ac{ofdm}-symbol index}
  \label{tab:symbols}
  \centering
  \footnotesize{
  \ifonecolumn 
  \else
  \renewcommand{\arraystretch}{1.3} 
  \fi
   \begin{tabular}{ccl}
     \toprule
     Notation               & Dim. & Description \\ 
     \midrule
     $M$                    & $\mathbb{N}$ & Number of subbands \\
     $N$                    & $\mathbb{N}$ & \acs{fc} processing \acs{ifft} length \\
     $L_m$                  & $\mathbb{N}$ & \acs{fc} processing \acs{fft} length \\
     $I_m$                  & $\mathbb{N}$ & \acs{fc} processing interpolation factor \\
     $B_{m}$                & $\mathbb{N}$ & Number of \acs{ofdm} symbols \\
     $R_m$                  & $\mathbb{N}$ & Number of \acs{fc} processing blocks  \\
     $L_{\text{act},m,n}$   & $\mathbb{N}$ & Number of active subcarriers ($=12L_{\text{PRB},m,n}$)\\
     $L_{\text{PRB},m,n}$   & $\mathbb{N}$ & Number of active \acsp{prb} ($=L_{\text{act},m,n}/12$)\\
     $L_{\text{CP},m,n}$    & $\mathbb{N}$ & Low-rate \acs{cp} length in samples\\
     $L_{\text{OFDM},m}$    & $\mathbb{N}$ & Low-rate \acs{ofdm} \acs{ifft} transform length\\
     $N_{\text{CP},m,n}$    & $\mathbb{N}$ & High-rate \acs{cp} length in samples\\
     $N_{\text{OFDM},m}$    & $\mathbb{N}$ & High-rate \acs{ofdm} \acs{ifft} transform length\\
     $f_\text{s}$           & $\mathbb{R}$ & Sampling frequency [Hz] \\
     $f_{\text{BS},m}$      & $\mathbb{R}$ & \acs{fc} processing bin spacing [Hz] \\
     $f_{\text{SCS},m,n}$   & $\mathbb{R}$ & \acs{ofdm} \acl{scs} [Hz] \\
     $N^\text{slot}_\text{subframe}$ & $\mathbb{N}$ & Number of slots per subframe \\
     $N^\text{symb}_\text{slot}$     & $\mathbb{N}$ & Number of symbols per slot \\
     $N^\text{samp}_\text{HSF}$      & $\mathbb{N}$ & Number of samples per half subframe \\   
     $f_\text{Ch,BW}$                & $\mathbb{R}$ & \ac{5gnr} or \ac{lte} channel bandwidth [Hz] \\
     \bottomrule 
    \end{tabular}}   
\end{table}
\fi

\begin{table*}[t!]   
  \caption{\acs{5g-nr} mixed numerology in frequency range 1 (FR1).}
  \label{tab:numerology2}
  \centering
  \begin{tabular}{lcccc}
    \toprule
    Subcarrier spacing, $f_{\text{SCS},m,n}$
    & \SI{15}{kHz} & \SI{30}{kHz} & \SI{60}{kHz} &  
    ${2^\mu\times}\SI{15}{kHz}$ \\
    \midrule
    OFDM symbol duration, $T_{\text{OFDM},m,n}$
    & \SI{66.7}{\mics}  & \SI{33.3}{\mics} & \SI{16.7}{\mics} & $2^{-\mu}\times\SI{66.6}{\mics}$ \\
    Cyclic prefix duration, $T_{\text{CP},m,n}$
    & \SI{4.69}{\mics}  & \SI{2.34}{\mics} & \SI{1.17}{\mics} & $2^{-\mu}\times\SI{4.69}{\mics}$ \\    
    Number of OFDM symbols per slot, $N^\text{symb}_\text{slot}$
    & 14                & 14               & 14 or 12         & 14 or 12 \\     
    Number of slots per subframe, $N^\text{slot}_\text{subframe}$
    & 1                 & 2                & 4                & $2^\mu$ \\
    Slot duration, $T_\text{slot}$
    & \SI{1}{ms}        & \SI{0.5}{ms}     & \SI{0.25}{ms}    & $2^{-\mu}\times\SI{1}{ms}$ \\
    \bottomrule
  \end{tabular} 
\end{table*}

\section{5G New Radio Scalable Numerology and Frame Structure}%
\label{sec:5Gnumerology}
\ac{cp-ofdm} has been selected as the baseline waveform for \ac{5gnr} at below \SI{52.6}{GHz} frequency bands, while \ac{dft-s-ofdm} (also known as \ac{sc-fdma}) can also be used for \ac{ul} in coverage-limited scenarios. \ac{5gnr} provides scalable numerology and frame structure in order to support diverse services, deployment scenarios, and user requirements operating from bands below \SI{1}{GHz} to bands above \SI{30}{GHz} known as \ac{mmwave}.  This numerology supports multiple \acp{scs} in order to reduce latency and to provide increased robustness to phase noise and Doppler, especially at higher carrier frequencies. In addition, by increasing the \ac{scs}, the maximum channel bandwidth supported for a given \ac{ofdm} transform length can be increased. On the other hand, smaller \acp{scs} have the benefit of providing longer \ac{cp} durations in time and, consequently, better tolerance to delay spread with reasonable overhead \cite{B:Dahlman2018}. These smaller \acp{scs} also allow transmitters to increase the power spectral density of the transmitted signal, which can be used for extending \ac{5g-nr} coverage.


\ac{5gnr} supports \acfp{scs} of $2^\mu\times\SI{15}{kHz}$ where $\mu=0,1,\ldots,4$ while only \SI{15}{kHz} is supported by \ac{lte} . Similar to \ac{lte} technology, a radio frame of \SI{10}{ms} is divided into \num{10} subframes, each having \SI{1}{ms} duration while each subframe has $2^{\mu}$ slots. Each slot consists of either~\num{14} or~\num{12} \ac{ofdm} symbols for the normal \ac{cp} or extended \ac{cp}, respectively \cite{S:3GPP:TS38.104v164,B:Dahlman2018}. The slot duration varies based on the \ac{scs} as $T_\text{slot}=2^{-\mu}\times\SI{1}{ms}$, i.e., it is \SI{1}{ms} for \SI{15}{kHz} \ac{scs}, \SI{0.5}{ms} for \SI{30}{kHz} \ac{scs} and so on.  The numerology for \SI{15}{kHz}, \SI{30}{kHz}, and \SI{60}{kHz} \ac{scs} is summarized in Table~\ref{tab:numerology2}.

\begin{figure}[t!]            
  \centering 
  \ifonecolumn 
    \includegraphics[width=0.65\textwidth]{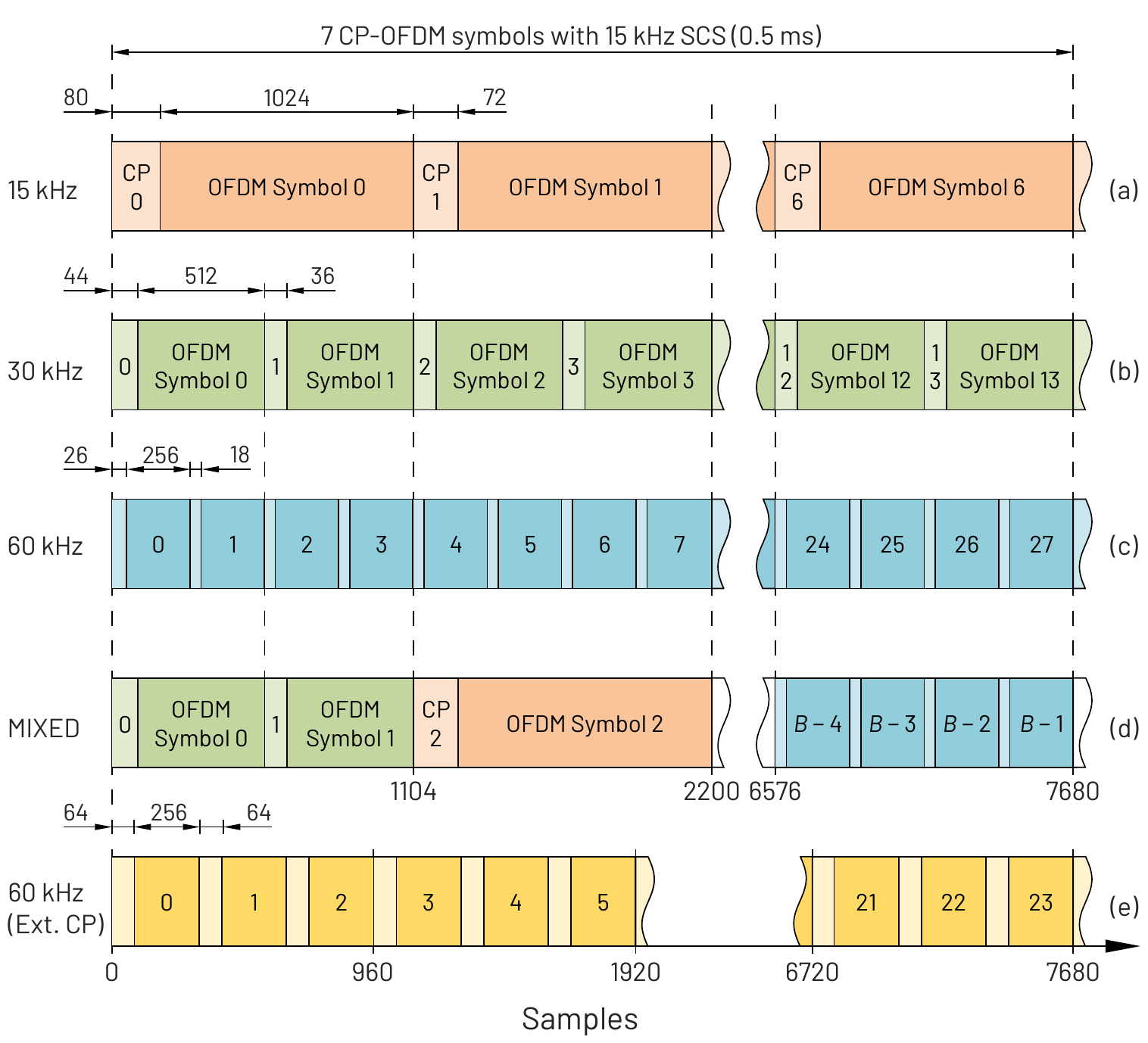}
  \else 
    \includegraphics[width=\figwidth]{Figs/MultiNumer2021Jun14.pdf} 
  \fi
  \caption{Illustration of multiple numerologies in \SI{10}{MHz} channel and their synchronization within the half subframe (\SI{0.5}{ms}). \mbox{(a)--(c)} \SI{15}{kHz}, \SI{30}{kHz}, and \SI{60}{kHz} \ac{scs} with normal \ac{cp}, respectively. (d) Time-multiplexed mixed numerology case. (e) \SI{60}{kHz} \ac{scs} with extended \ac{cp}.
  } 
  \label{fig:Numerology}  
\end{figure} 

Fig.~\ref{fig:Numerology} exemplifies the time alignment of different numerologies within a half subframe. In general, \ac{5gnr} does not specify the minimum number of consecutive symbols of certain \ac{scs} and, therefore, in extreme time-multiplexing cases, it is possible that the \ac{scs} changes even at the symbol level as illustrated in Fig.~\ref{fig:Numerology}(d). 

\ac{5gnr} supports also non-slot based scheduling (so-called mini slots), where the transmission length can be configured between~\num{1} and~\num{13} symbols  \cite[Section 8.1]{R:3GPP:TR38.912v150}. This mini-slot concept is especially essential for low-latency \ac{urllc} and for dynamic time-domain multiplexing. Such transmissions can pre-empt the ongoing slot-based transmission and, therefore, also the time-domain flexibility of the filtering solutions becomes crucial in the \acs{5g-nr} context.

\ac{5gnr} allows the use of a mixed numerology, i.e., using different \acp{scs} in different subbands (or \acp{bwp}) within a single channel. However, the use of different \acp{scs} within an \ac{ofdm} multiplex harms the orthogonality of subcarriers, introducing \ac{ini}. To cope with \ac{ini} in mixed-numerology scenarios with basic \ac{cp-ofdm} waveform, relatively wide \acp{gb} should be applied between adjacent \acp{bwp}, which would reduce the spectral efficiency. Alternatively, the \ac{3gpp} allows to use spectrum enhancement techniques for \ac{cp-ofdm}, but this should be done in a transparent way and without performance loss with respect to plain \ac{cp-ofdm}.  The transparency means that a \ac{tx} or a \ac{rx} does not need to know whether spectrum enhancement is used at the other end. The spectrum enhancement techniques should also be compatible with each other, allowing different techniques to be used in the \ac{tx} and \ac{rx}. Transparent enhanced \ac{cp-ofdm} techniques have been considered in~\cite{C:Bazzi_EuCNC2017, J:Zayani_Access2018, J:Levanen_WC2019}.

\ac{5gnr} is designed to operate in two operating bands where \ac{fr1} corresponds to \SIrange{410}{7125}{MHz} while \ac{fr2} corresponds to the \SIrange{24.25}{52.6}{GHz} \cite[Table 5.1-1]{S:3GPP:TS36.104v166}.
Basically, \ac{fr1} supports \acp{scs} of $f_\text{SCS}=2^\mu\times\SI{15}{kHz}$ with $\mu=\{0,1,2\}$ and  $\mu=\{2,3,4\}$ for \ac{fr2}. \ac{scs} of $f_\text{SCS}=2^\mu\times\SI{15}{kHz}$ with $\mu=\{0,1,3,4\}$ are used for synchronization blocks (\ac{pss}, \ac{sss}, and \ac{pbch}) and $\mu=\{0,1,2,3\}$ for data channels (\ac{pdsch}, \ac{pusch}, etc.). 

For the approaches proposed in this manuscript, \emph{the \ac{scs} for each \ac{cp-ofdm} symbol on each subband can be independently adjusted.} Therefore, we denote the \ac{scs} for the $n$th symbol on the $m$th subband by $f_{\text{SCS},m,n}$ for $n\in\{0,1,\dots,B_m-1\}$ and $m\in\{0,1,\dots,M-1\}$, where $B_m$ and $M$ are the number of \ac{cp-ofdm} symbols on subband $m$ and number of subbands, respectively. Here, $f_{\text{SCS},m,n}$ can be selected as $2^\mu\times\SI{15}{kHz}$ for $\mu=0,1,\dots,4$ and the \ac{scs} scaling factor of $n$th symbol on subband $m$ is denoted by $\mu_{m,n}$.

\ifonecolumn 
\begin{table}[t!]   
  \centering
  \medskip
  \caption{Nominal channel bandwidths ($f_\mathrm{Ch,BW}$) for \acs{5g-nr} FR1 and corresponding sample rates ($f_\mathrm{s}$)}
  \label{tab:5gnr_channels_fr1}
  \addtolength{\tabcolsep}{-1.0pt}
  \begin{tabular}{r*{13}{S[table-format=2.2]S[table-format=2.2]}}
    \toprule
    $f_\text{Ch,BW}$ [MHz] & 5.00 & 10.00 & 15.00 & 20.00 & 25.00 & 30.00 & 40.00 & 50.00 & 60.00 & 70.00 &  80.00 &  90.00 & 100.00 \\
    $f_\text{s}$ [MHz]  & 7.68 & 15.36 & 23.04 & 30.72 & 30.72 & 46.08 & 61.44  & 61.44 & 92.16 & 92.16 & 122.88 & 122.88 & 122.88 \\
    \bottomrule
  \end{tabular}
\end{table}
\else
\begin{table}[t!]   
  \centering
  \medskip
  \caption{Nominal channel bandwidths ($f_\mathrm{Ch,BW}$) for \acs{5g-nr} frequency range 1 (FR1) and corresponding sample rates ($f_\mathrm{s}$)}
  \label{tab:5gnr_channels_fr1}
  \addtolength{\tabcolsep}{-1.0pt}
  \begin{tabular}{r*{13}{S[table-format=2.2]S[table-format=2.2]}}
    \toprule
    $f_\text{Ch,BW}$ [MHz] & 5.00 & 10.00 & 15.00 & 20.00 & 25.00 & 30.00 & 40.00 \\
    $f_\text{s}$ [MHz]  & 7.68 & 15.36 & 23.04 & 30.72 & 30.72 & 46.08 & 61.44 \\
    \midrule[0.75pt]
    $f_\text{Ch,BW}$ [MHz] & 50.00 & 60.00 & 70.00 &  80.00 &  90.00 & 100.00 \\
    $f_\text{s}$ [MHz]  & 61.44 & 92.16 & 92.16 & 122.88 & 122.88 & 122.88 \\
    \bottomrule
  \end{tabular}
\end{table}
\fi

Let $f_\text{s}$ be the \ac{ofdm} waveform sample rate as tabulated in Table~\ref{tab:5gnr_channels_fr1} for \ac{5g-nr} channel bandwidths in \ac{fr1}. Without loss of generality, we assume that the number of samples to be processed is multiple of
\begin{equation}
  N^\text{samp}_\text{HSF} = 0.5\times10^{-3}f_\text{s},
\end{equation}
i.e., number of samples per half subframe corresponding to seven \ac{cp}-\ac{ofdm} symbols with baseline \ac{scs}. The baseline \ac{scs} is  $f_\text{BL}=\SI{15}{kHz}$ and $f_\text{BL}=\SI{60}{kHz}$ in \ac{fr1} and \ac{fr2}, respectively. The \ac{ofdm} transform length can now be determined as a ratio of sample rate and \ac{scs} as
\begin{equation} 
  N_{\text{OFDM},m,n} =
  \frac{f_\text{s}}{f_{\text{SCS},m,n}}=
  \frac{f_\text{s}}{2^{\mu_{m,n}}\times\SI{15}{kHz}}.
\end{equation}
The maximum available \ac{ofdm} \ac{ifft} length in \ac{5gnr} is restricted to be smaller than or equal to $N_\text{IFFT,max}=4096$, therefore, the maximum supported channel bandwidth, e.g., for $\SI{15}{kHz}$ \ac{scs} is $\SI{50}{MHz}$ while $\SI{240}{kHz}$ \ac{scs} supports $\SI{800}{MHz}$.

The normal \ac{cp} length in samples is determined as
\begin{subequations}
  \begin{equation}
    N_{\text{CP},m,n} =
    \begin{dcases}
      \frac{9}{128}N_{\text{OFDM},m,n} + \alpha, &
      \text{for $\xi(n)=0$} \\
      \frac{9}{128}N_{\text{OFDM},m,n}, &
      \text{otherwise},
  \end{dcases}    
  \end{equation}
  where
  \begin{equation}
    \label{eq:alpha}
    \alpha = \bmod\left(N^\text{samp}_\text{HSF}, 9+128\right)
  \end{equation}
  and 
  \begin{equation}
    \xi(n) = \bmod\left(
      N_{\text{OFDM},m,0}-
      \sum_{k=0}^{n} 
      N_{\text{OFDM},m,k}, \frac{7f_\text{s}}{f_\text{BL}}
    \right)    
  \end{equation}
\end{subequations}
for $n=0,1,\dots,B_m-1$ is equal to zero for the first symbol of each half subframe. In \ac{5g-nr} numerology (as well as in \ac{lte}), longer \ac{cp} for the first symbol is needed to balance the excess samples for each half subframe such that
\begin{equation}
  \sum_{n=0}^{B_m-1}\left(N_{\text{OFDM},m,n}+N_{\text{CP},m,n}\right)=N^\text{samp}_\text{HSF}
\end{equation}
for given number of \ac{cp-ofdm} symbols $B_m$. For example, in \SI{10}{MHz} channel ($f_\text{s}=\SI{15.36}{MHz}$) with \SI{15}{kHz} \ac{scs} and seven ($B_m=7$) \ac{ofdm} symbols per half subframe, $\alpha=8$ and the \ac{cp} length is $N_{\text{CP},m,n}=72+\alpha=80$ for $\bmod(n,7)=0$ and $N_{\text{CP},m,n}=72$ otherwise, such that $7(1024+72)+8=N_\text{HSF}^\text{samp}$.
 
In \ac{lte} and \ac{5gnr}, the frequency-domain resources are allocated in \acp{prb} corresponding to \num{12} subcarriers or \acp{re}. The \emph{transmission bandwidth configuration} defining the maximum number of active \acp{prb} for given channel bandwidth and given \ac{scs} are tabulated in \cite[Tables 5.3.2-1 and 5.3.2-2]{S:3GPP:TS36.104v166} for \acs{fr1} and \acs{fr2}, respectively.

The following processing model supports mixed \acp{scs} and allocation bandwidths. Therefore, we denote by $L_{\text{act},m,n}$, the number of active subcarriers of the $n$th symbol on subband $m$ and $\mathcal{S}_{m,\upsilon}\subset\{0,1,\dots,B_{m}-1\}$ for $\upsilon=0,1,\dots,\Upsilon_m-1$ is the set of symbol indices having the same symbol length and the same number of active subcarriers while $\Upsilon_m$ is the number of symbol sets with different numerology on subband $m$. Here, the number of active subcarriers can be selected as $L_{\text{act},m,n}\leq 12\times N_{\text{PRB,max}}$, where $N_{\text{PRB,max}}$ is the corresponding transmission bandwidth configuration. 

\begin{figure}[t!]            
  \centering
  \ifonecolumn 
    \includegraphics[trim=20 0 50 0,clip,
    width=0.65\textwidth]{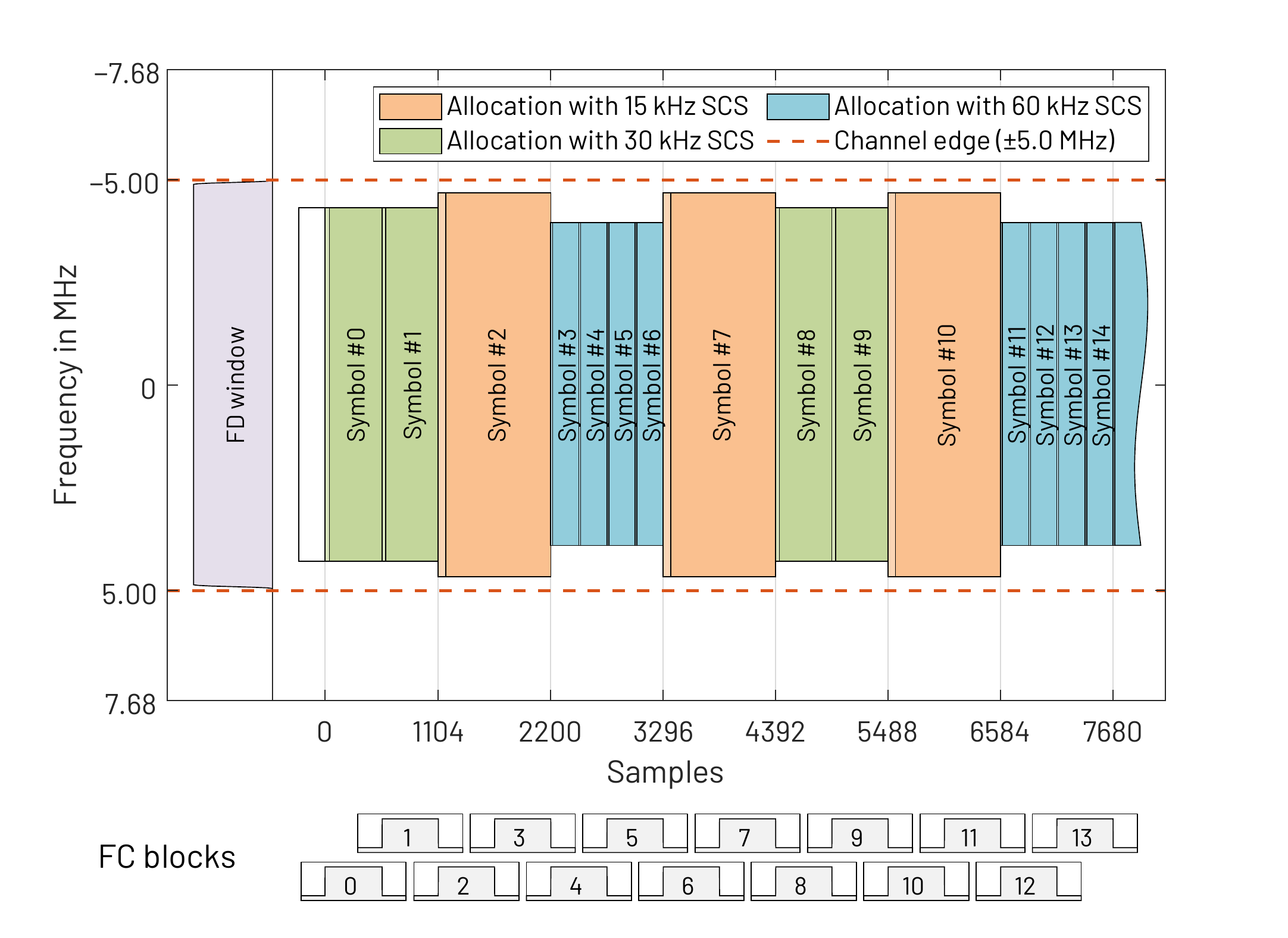} 
  \else
    \includegraphics[trim=20 0 50 0,clip,width=\columnwidth]{Figs/TFA_M2_10MHz.pdf}
  \fi
  \caption{Time-multiplexed mixed-numerology scenario with three \acp{scs}. In this example, $\mathcal{S}_{0,0}=\{2,7,10\}$ is the set of symbol indices with \SI{15}{kHz} \ac{scs}, $\mathcal{S}_{0,1}=\{0,1,8,9\}$ are the indices for symbols with \SI{30}{kHz} \ac{scs}, and $\mathcal{S}_{0,2}=\{3,4,5,6,11,12,13,14\}$ for \SI{60}{kHz} \ac{scs}. The number of active subcarriers in \SI{10}{MHz} channel for symbols with \SI{15}{kHz}, \SI{30}{kHz}, and \SI{60}{kHz} \acp{scs} are \num{624}, \num{288}, and \num{132}, respectively \cite{S:3GPP:TS38.104v164}.}
    \label{fig:MixedBlock}      
\end{figure}  

Fig.~\ref{fig:MixedBlock} illustrates a single subchannel ($M=1$) time-multiplexed mixed-numerology scenario with three ($\Upsilon_0=3$) \acp{scs}. In this example, $\mathcal{S}_{0,0}=\{2,7,10\}$ is the set of symbol indices with \SI{15}{kHz} \ac{scs} while $\mathcal{S}_{0,1}=\{0,1,8,9\}$ and $\mathcal{S}_{0,2}=\{3,4,5,6,11,12,13,14\}$ are the indices for symbols with \SI{30}{kHz} and \SI{60}{kHz} \ac{scs}, respectively.  In this scenario, we consider \SI{10}{MHz} \ac{5g-nr} channel, where \ac{ofdm} transform lengths are $N_{\text{OFDM},m,n}=1024$ for $n\in\mathcal{S}_{0,0}$, $N_{\text{OFDM},m,n}=512$ for $n\in\mathcal{S}_{0,1}$, and $N_{\text{OFDM},m,n}=256$ for $n\in\mathcal{S}_{0,2}$ while the corresponding number of active subcarriers are $L_{\text{act},m,n}=12\times52=\num{624}$, $L_{\text{act},m,n}=12\times24=\num{288}$, and $L_{\text{act},m,n}=12\times11=\num{132}$, respectively. The sampling rate is $f_{\text{s}}=\SI{15.36}{MHz}$ corresponding  to $N^\text{samp}_\text{HSF}=\num{7680}$ samples within the half subframe. 
 
\section{Fast-Convolution-based Filtered-OFDM}
\begin{figure}[t!]               
  \centering    
  \ifonecolumn 
    \includegraphics[width=0.6\textwidth]{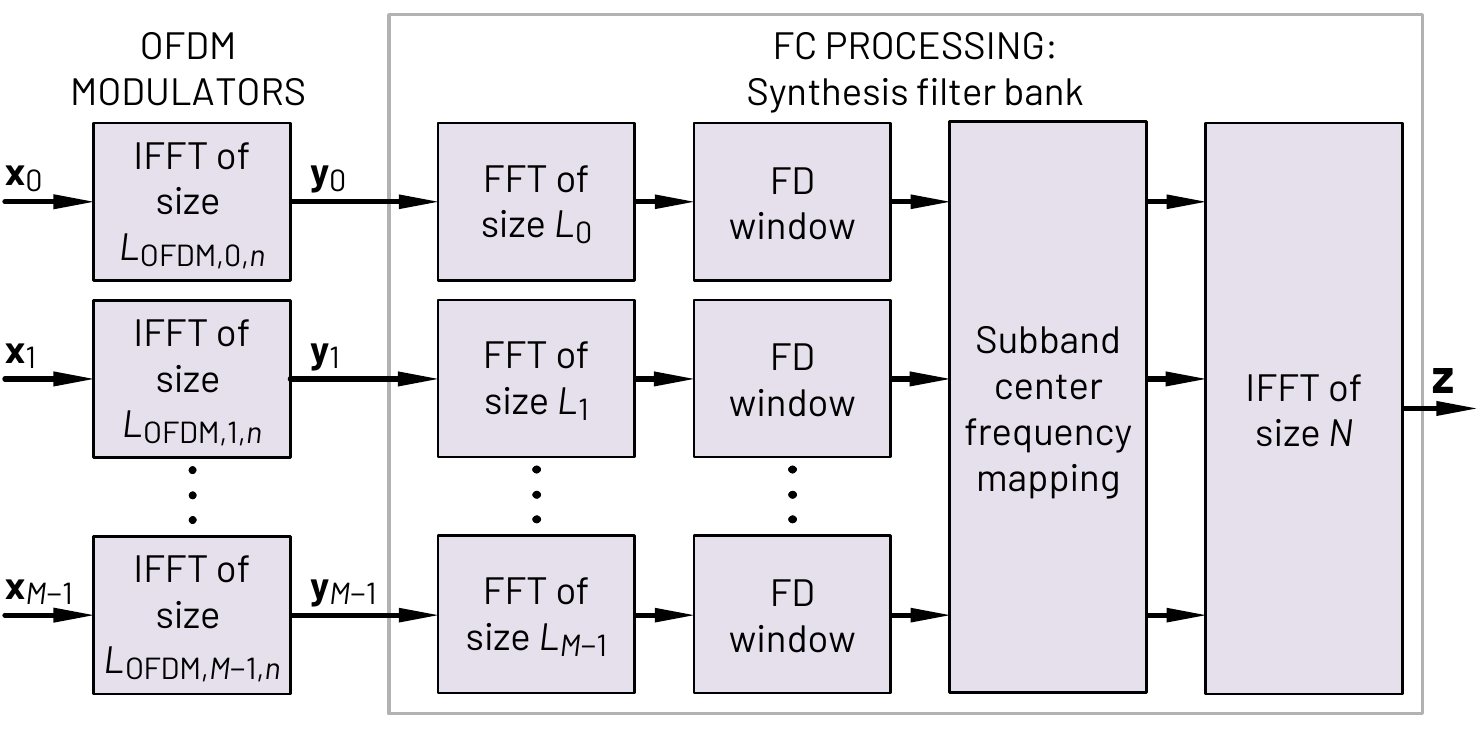}
  \else
    \includegraphics[width=0.995\columnwidth]{Figs/FC-TX2021Nov17.pdf}
  \fi
  \caption{Proposed transmitter processing using the \acs{fc}
    \acf{sfb} with $M$ subbands. In \acs{fc}-based filtered-\acs{ofdm}, filtering is applied at subband level, which means one or multiple contiguous \acsp{prb} with possible mixed \acs{scs}, while utilizing normal CP-OFDM waveform for the \acsp{prb}. \acs{fc}-processing consist of forward transforms of length $L_m$ for $m=0,1,\dots,M$ and inverse transform of length $N$. The center frequency of each symbol may be adjusted independently by simply mapping the corresponding frequency-domain bins.}  
  \label{fig:FCTX}         
\end{figure}

\acused{fc}
\label{sec:fc-f-ofdm}
The basic principle of the proposed \ac{fc}-based waveform \ac{tx} processing for \ac{5gnr} is illustrated in Fig.~\ref{fig:FCTX}. In original \ac{fc-f-ofdm}, filtering is applied at subband level, utilizing normal \ac{cp}-\ac{ofdm} waveform with one or multiple contiguous \acp{prb} with same \ac{scs} \cite{C:Renfors2015:fc-f-ofdm, C:Renfors16:adjustableCP, J:Yli-Kaakinen:JSAC2017,C:Yli-Kaakinen:ASILOMAR2018,J:Yli-Kaakinen:TWC2021}. In the proposed model, each subband can have mixed numerology, that is, \ac{scs} and/or number of active \acp{prb} may change from one \ac{ofdm} symbol to another. These \ac{ofdm} symbols are generated with \acp{ifft} of length $L_{\text{OFDM},m,n}$ for $m=0,1,\dots,M-1$ and $n=0,1,\dots,B_m-1$. \ac{cp} of length $L_{\text{CP},m,n}$ is inserted to each symbol and the resulting \ac{cp-ofdm} symbols are filtered using \ac{fc}-based \ac{sfb} consisting of forward transforms (\acp{fft}) of length $L_m$ for $m=0,1,\dots,M-1$, frequency-domain windowing, and inverse transform (\ac{ifft}) of length $N$. The center frequency of each subband can be adjusted simply by mapping the windowed output bins of the forward transforms (FFTs) to the desired input bins of the inverse transform (IFFT).

The \ac{fc} bin spacing, i.e, the resolution of the \ac{fc} processing, is determined as a ratio of output sample rate and \ac{fc} inverse transform length as
\begin{equation}
  \label{eq:BS}
  f_\text{BS} = f_\text{s}/N.
\end{equation}
The \ac{fc} bin spacing can be selected independent to \ac{ofdm} \ac{scs}. \ac{fc} processing provides the sampling-rate conversion factor determined by the ratio of the inverse transform and forward transform length as expressed by
\begin{equation} 
  I_m = N/L_m. 
\end{equation}
Therefore, the \ac{ofdm} symbol and \ac{cp} lengths on the high-rate side (at the \ac{sfb} output) are $N_{\text{OFDM},m,n}=I_m L_{\text{OFDM},m,n}$ and $N_{\text{CP},m,n}=I_m L_{\text{CP},m,n}$, respectively. By following the proposed segmentation of the subband waveforms into the overlapping \ac{fc} blocks and then carrying out the overlapped circular convolutions with the aid of \ac{fc}-based \ac{sfb} (\ac{fft}/\ac{ifft} pair with windowing) in conjunction of \ac{ola} or \ac{ols} schemes, even the center frequency of each symbol may adjusted independently.

\subsection{FC Filtered-OFDM TX Processing}
Let us denote the $n$th frequency-domain multi-carrier symbol on $m$th subband with $L_{\text{act},m,n}$ active subcarriers by
\begin{equation}
  \mathbf{x}_{m,n}=
 \begin{bmatrix}
    x_{m,n}(0) &
    x_{m,n}(1) &
    \dots &
    x_{m,n}(L_{\text{act},m,n}-1) 
  \end{bmatrix}^\transpose.
\end{equation}
Here, $x_{m,n}(\ell)$'s are the \ac{qpsk} or $\mathcal{M}$-ary \acl{qam} ($\mathcal{M}$-\acs{qam})\acused{qam} symbols to be transmitted on subcarrier $\ell$.
Further,
\begin{subequations}  
\begin{equation}
  \mathbf{x}_m=
  \begin{bmatrix}
    \mathbf{x}_{m,0}^\transpose &
    \mathbf{x}_{m,1}^\transpose &
    \dots &
    \mathbf{x}_{m,B_{m}-1}^\transpose 
  \end{bmatrix}^\transpose
\end{equation} 
is the column vector of length
\begin{equation}
  L_{\text{act,tot},m}=
  \sum_{n=0}^{B_{m}-1}L_{\text{act},m,n}
\end{equation}
\end{subequations}
obtained by vertically stacking all symbols $\mathbf{x}_{m,n}$ for $n=0,1,\dots,B_{m}-1$. The corresponding low-rate \ac{cp-ofdm} waveform length is
\begin{equation}
    L_{\text{samp},m} = 
    \sum_{n=0}^{B_{m}-1}\left(
    L_{\text{OFDM},m,n}+L_{\text{CP},m,n}
    \right).    
\end{equation}

\begin{figure}[t!]             
  \centering   
  \ifonecolumn 
    \includegraphics[width=0.6\columnwidth]{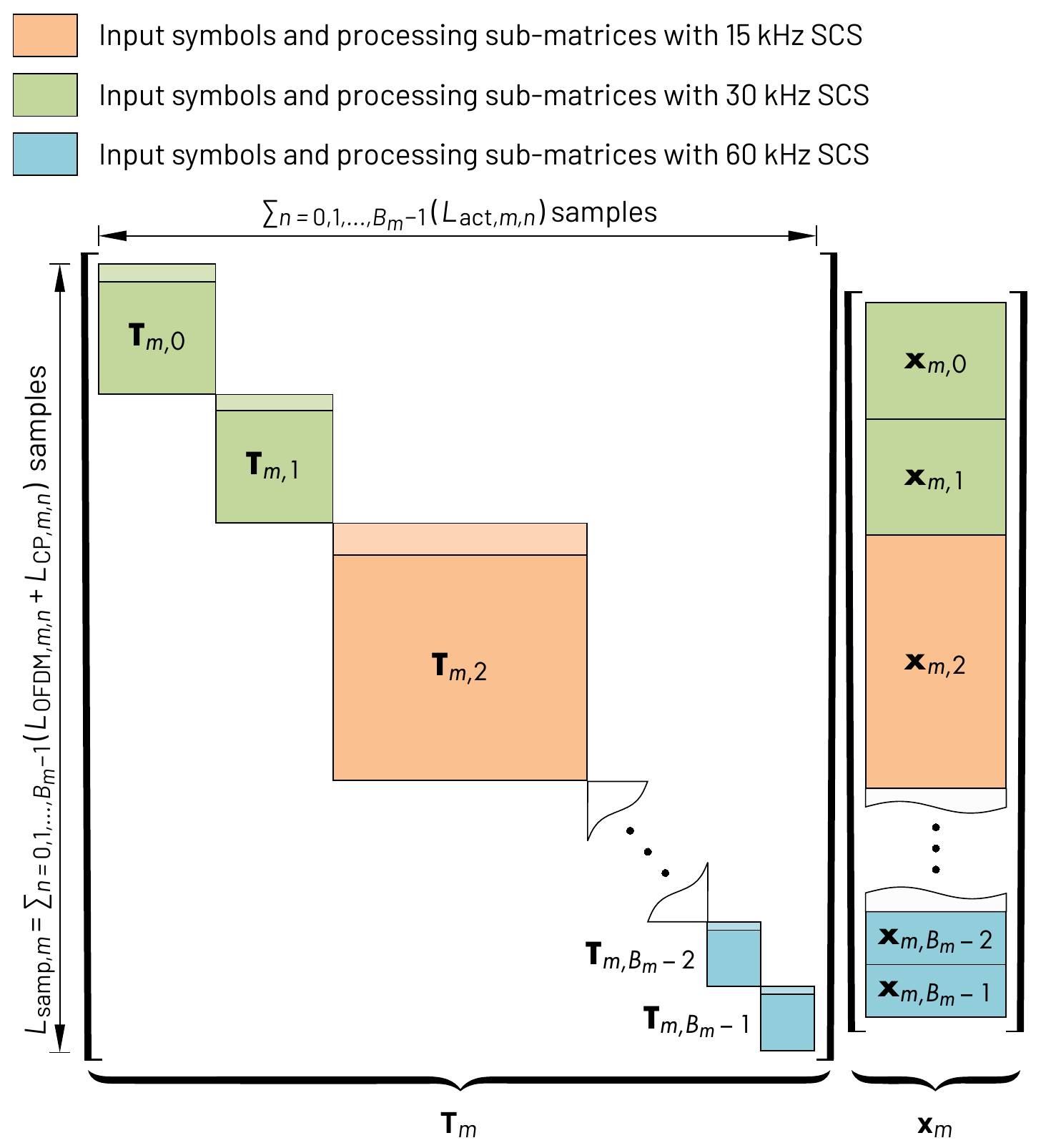}
  \else
    \includegraphics[width=0.95\columnwidth]{Figs/OFDMProcessingMatrix2021Jul22.pdf}
  \fi
  \caption{Illustration of the structure of block-diagonal \ac{ofdm} modulation matrix $\mathbf{T}_m$ with $B_m$ blocks.}   
  \label{fig:OFDMmodMtx}         
\end{figure}

The \ac{cp}-\ac{ofdm} \ac{tx} processing of the $m$th subband can now be expressed as
\begin{subequations}  
  \begin{equation}
    \label{eq:y_vect}
    \mathbf{y}_{m}=
    \begin{bmatrix}
      \mathbf{y}_{m,0}^\transpose &
      \mathbf{y}_{m,1}^\transpose &
      \dots &
      \mathbf{y}_{m,B_{m}-1}^\transpose 
    \end{bmatrix}^\transpose =
    \mathbf{T}_m\mathbf{x}_m,
  \end{equation}
  where the block diagonal \ac{ofdm} modulation matrix of size $L_{\text{samp},m}\times L_{\text{act,tot},m}$ is expressed as
  \begin{equation}
    \mathbf{T}_m = 
    \diag*{
      \mathbf{T}_{m,0},\mathbf{T}_{m,1},    
      \dots,\mathbf{T}_{m,B_{m}-1}
    }
  \end{equation}
  with 
  \begin{equation}
  \label{eq:bwOFDMmod}
    \mathbf{T}_{m,n}=
    \mathbf{K}_{m,n}
    \widehat{\mathbf{W}}_{m,n}^\mathsf{H}.
  \end{equation}
\end{subequations}
Here, $\widehat{\mathbf{W}}_{m,n}$ is  pruned unitary \ac{dft} matrix of size $L_{\text{act},m,n}\times L_{\text{OFDM},m,n}$ as given by
\begin{equation} 
  \label{eq:prunedDFT} 
  \left[\widehat{\mathbf{W}}_{m,n}\right]_{p,q}= 
  \frac{1}{\sqrt{L_{\text{OFDM},m,n}}}
  \exp\left(\frac{
  -\iu\pi q\left(2p-L_{\text{act},m,n}\right)
  }{L_{\text{OFDM},m,n}}\right)
\end{equation}
for $p=0,1,\dots,L_{\text{act},m,n}-1$ and $q=0,1,\dots,L_{\text{OFDM},m,n}-1$.  In \eqref{eq:bwOFDMmod}, $\mathbf{K}_{m,n}\in\mathbb N^{(L_{\text{OFDM},m,n}+L_{\text{CP},m,n})\times L_{\text{OFDM},m,n}}$ is the \ac{cp} insertion matrix as given by
\begin{equation}
  \mathbf{K}_{m,n} =
  \begin{bmatrix}
    \begin{matrix}
      \mathbf{0}_{L_{\text{CP},m,n}\times(L_{\text{OFDM},m,n}-L_{\text{CP},m,n})} &
      \mathbf{I}_{L_{\text{CP},m,n}} \\
    \end{matrix}\\
    \mathbf{I}_{L_{\text{OFDM},m,n}}
  \end{bmatrix}
\end{equation}
which copies $L_{\text{CP},m,n}$ last samples of the $n$th \ac{ofdm} symbol in the beginning of the symbol. The block diagonal structure of the resulting \ac{ofdm} modulation matrix $\mathbf{T}_{m}$ is illustrated in Fig.~\ref{fig:OFDMmodMtx}. 

\begin{figure}[t!]            
  \centering 
  \ifonecolumn 
    \includegraphics[width=0.67\columnwidth]{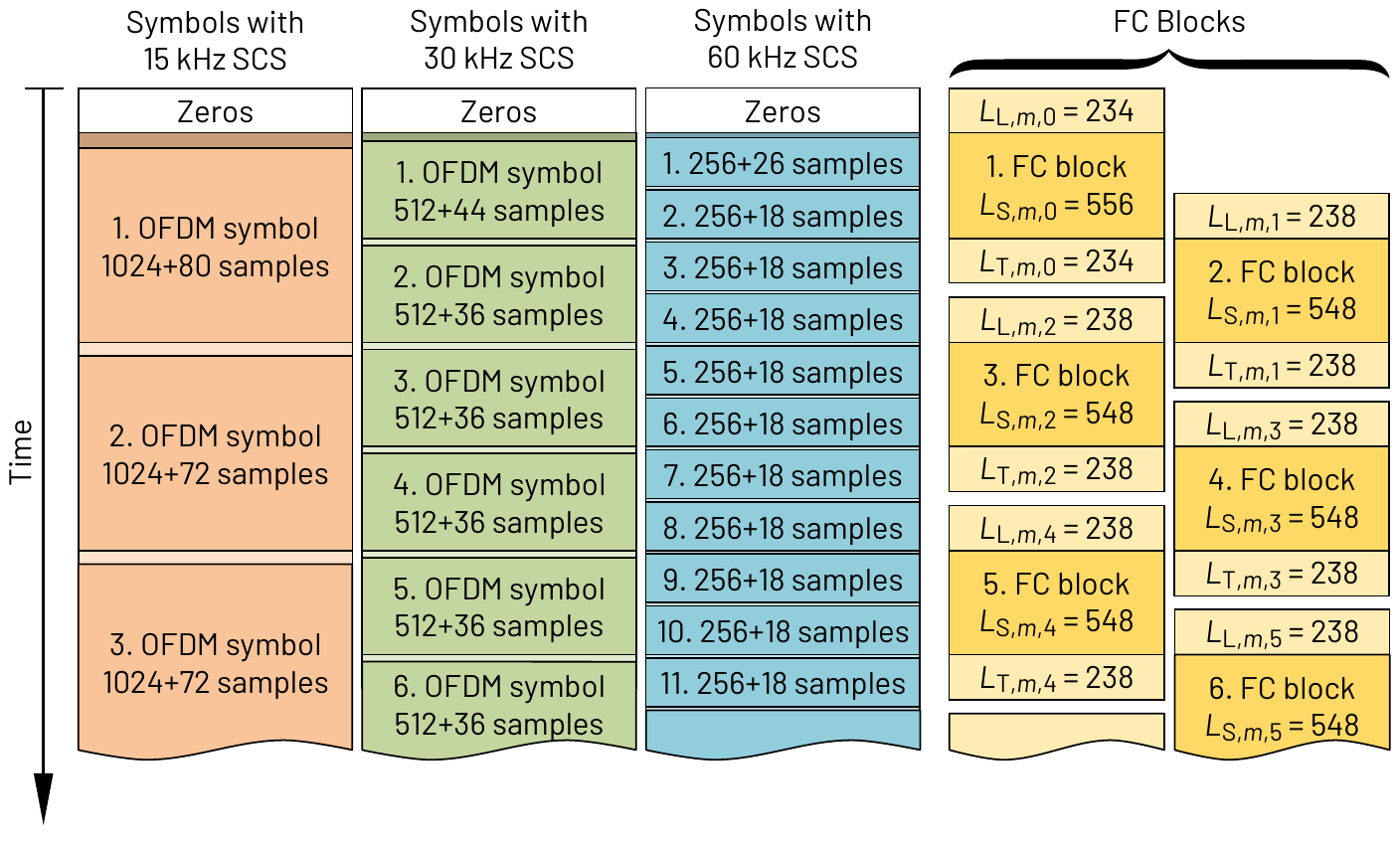}  
  \else 
    \includegraphics[width=\figwidth]{Figs/SymbSynchCont_L1024_ex.pdf} 
  \fi
  \caption{Illustration of the block segmentation in continuous symbol-synchronized \ac{fc} processing. By adjusting the non-overlapping part length $L_{\text{S},m,n}$, the \ac{fc} blocks can be synchronized with all numerologies.  For first \ac{fc}-processing block of each half subframe the non-overlapping part length is larger. 
  }     
  \label{fig:FCblocking}  
\end{figure}

The block segmentation of the proposed continuous symbol-synchronized \ac{fc}-processing scheme is exemplified in Fig.~\ref{fig:FCblocking}. Time-domain input sample stream is processed in overlapped \ac{fc}-processing blocks of length $L_m$ as in earlier schemes. However, now the overlap between the processing blocks is adjusted such that the length of the non-overlapping part for the \ac{fc} blocks containing the \ac{cp} part of the first symbol in a half subframe is longer than others.

Let us denote by $R_\text{HSF}$ the number of half-subframes to be processed. For the proposed scheme, the non-overlapping part length for \ac{fc} blocks for $r=0,1,\dots,R_\text{HSF}R_m-1$ is given by
\begin{subequations}
\begin{equation} 
  L_{\text{S},m,r}=
  \begin{cases}
    2^{\beta_m}(9+128) + \alpha, & \text{for $\bmod(r, R_m) = 0$}\\
    2^{\beta_m}(9+128),          & \text{otherwise},\\
  \end{cases}
\end{equation}
where
\begin{equation} 
  \beta_m = \left\lfloor\frac{1}{2}\frac{L_m}{128}\right\rfloor
\end{equation}
\end{subequations}
and $\alpha$ is given by \eqref{eq:alpha}.
Here, the number of \ac{fc}-processing blocks within the half subframe is determined as
\begin{equation} 
  R_m =
  \begin{cases}
    14 & \text{for \SI{15}{kHz} \ac{fc} bin spacing} \\
    28 & \text{for \SI{30}{kHz} \ac{fc} bin spacing} \\
    56 & \text{for \SI{60}{kHz} \ac{fc} bin spacing}.
  \end{cases}
\end{equation}
For example, for \SI{15}{kHz} \ac{fc}-processing bin spacing in \SI{10}{MHz} channel, as illustrated in Fig.~\ref{fig:FCblocking}, \ac{fc}-processing forward transform length is $L_m=1024$. In this case, the non-overlapping part length for the first \ac{fc} block of each subframe is $L_{\text{S},m,r} = 556$ samples whereas for the other \ac{fc} blocks the corresponding number is  $L_{\text{S},m,r} = 548$ samples. The leading and tailing overlapping part lengths for \ac{fc} blocks are given as
\ifonecolumn
\begin{subequations}
  \begin{equation} 
    L_{\text{L},m,n}=\lceil (L_m-L_{\text{S},m,n})/2 \rceil 
    \qquad\text{and}\qquad 
    L_{\text{T},m,n}=L_m-L_{\text{L},m,n},
  \end{equation}
\end{subequations}
\else
\begin{subequations}
  \begin{equation} 
    L_{\text{L},m,n}=\lceil (L_m-L_{\text{S},m,n})/2 \rceil 
  \end{equation}
  and 
  \begin{equation} 
    L_{\text{T},m,n}=L_m-L_{\text{L},m,n},
  \end{equation}
\end{subequations}
\fi
respectively. In this case, exactly two \ac{fc}-processing blocks are needed to process one, two, or four \ac{ofdm} symbols with \SI{15}{kHz}, \SI{30}{kHz}, or \SI{60}{kHz} \ac{scs}, respectively, as shown in Fig.~\ref{fig:FCblocking}. For $L_m=512$, the corresponding number of \ac{fc}-processing blocks is four, that is, the filtering can be re-configured with the shortest (\SI{60}{kHz} \ac{scs}) \ac{ofdm} symbol time resolution. 

\begin{figure}[t!]             
  \centering   
  \ifonecolumn 
  \includegraphics[width=0.6\columnwidth]{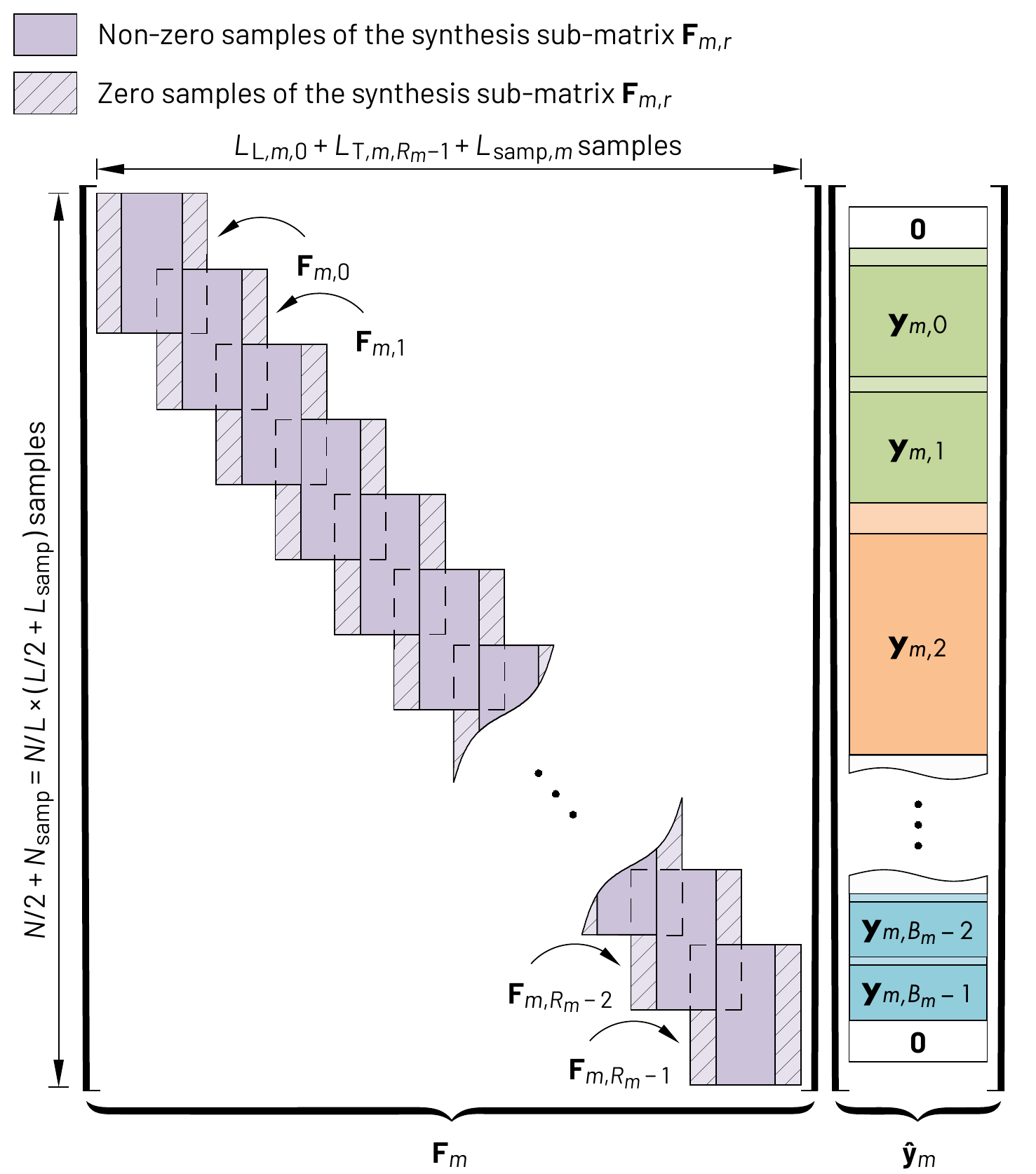}
  \else
    \includegraphics[width=0.95\columnwidth]{Figs/FCProcessingMatrix2021Jul23.pdf}
  \fi
  \caption{Illustration of the structure of \ac{fc} processing matrix $\mathbf{F}_m$ with $R_m$ overlapping blocks.}   
  \label{fig:FCprocessMtx}      
\end{figure}

In the \ac{fc} \ac{sfb} case, the block processing of $m$th \ac{cp}-\ac{ofdm} subband signal $\mathbf{y}_{m}$ for the generation of high-rate subband waveform $\mathbf{z}_m$ can now be represented as
\begin{subequations} 
  \label{eq:BDM}   
  \begin{equation} 
    \mathbf{z}_m = \mathbf{F}_m \hat{\mathbf{y}}_{m},
  \end{equation}
  where $\mathbf{F}_m$ is the block diagonal synthesis processing matrix of the form
  \begin{align}
    \label{eq:tx_matrix1}
    \mathbf{F}_m &= 
    \bdiag*{
      \mathbf{F}_{m,0},
      \mathbf{F}_{m,1},
      \dots,
      \mathbf{F}_{m,R_m-1}
    }_{q_{m,r},p_{m,r}}
  \end{align} 
with overlapping blocks $\mathbf{F}_{m,r}\in\mathbb C^{N\times L_m}$ for $r=0,1,\dots,R_m-1$ and
\begin{equation}
 \hat{\mathbf{y}}_{m}=
 \begin{bmatrix}
    \mathbf{0}_{L_{\text{L},m,0}\times 1}\\
    \mathbf{y}_{m}\\
    \mathbf{0}_{L_{\text{T},m,R_m-1}\times 1} 
  \end{bmatrix},
\end{equation}
\end{subequations}
is \ac{cp-ofdm} waveform of \eqref{eq:y_vect} with $L_{\text{L},m,0}$ and $L_{\text{T},m,R_m-1}$ samples zero padding before and after the \ac{cp-ofdm} symbols, respectively. Here, $\bdiag{\cdot}_{q_r,p_r}$ is an operator for constructing block-diagonal matrix with overlapping blocks of its arguments. The column and row indices of the first element of the $r$th block are $q_r$ and $p_r$, respectively. In order to align the \ac{fc}-processing blocks with \ac{cp-ofdm} symbols, the row and column indices for the first elements of the $r$th block are given as  
\begin{subequations}
\begin{equation}
  p_{m,r} = 
  \begin{dcases}
    r L_{\text{S},m,n} + \frac{\alpha}{2}\left\lfloor \frac{r}{R_m} \right\rfloor, & \text{for $\bmod(r,R_m)=0$} \\
    r L_{\text{S},m,n} + \frac{\alpha}{2}\Bigl\lceil\frac{r}{R_m}\Bigl\rceil, & \text{otherwise} 
  \end{dcases}
\end{equation}
and
\begin{equation}
  q_{m,r} = I_m p_{m,r},
\end{equation}
\end{subequations}
respectively. The block diagonal structure of $\mathbf{F}_m$ with overlapping blocks is depicted in Fig.~\ref{fig:FCprocessMtx}.

The overall waveform to be transmitted is obtained by summing all the $M$ subband waveforms as 
\begin{equation}
  \mathbf{z}=\sum_{m=0}^{M-1}\mathbf{z}_m. 
\end{equation}

\ac{fc} \ac{sfb} can be represented using block processing by decomposing the $\mathbf{F}_{m,r}$'s as follows:
\label{eq:synth}
\begin{equation}
  \label{eq:synthesis}
  \mathbf{F}_{m,r} =
  \mathbf{S}_r
  \mathbf{W}^\mathsf{H}_{N}
  \mathbf{M}_{m,r}
  \mathbf{D}_{m,r}
  \mathbf{P}_{L_m}
  \mathbf{W}_{L_m}
  \mathbf{A}_{m,r},
\end{equation}
where $\mathbf{A}_{m,r}=\diag{\mathbf{a}_{m,r}}$ and $\mathbf{S}_r=\diag{\mathbf{s}_r}$ are the time-domain analysis and synthesis windowing matrices with the analysis and synthesis window weights $\mathbf{a}_{m,r}\in\mathbb R^{L_m\times 1}$ and $\mathbf{s}_r\in\mathbb R^{N\times 1}$, respectively.  $\mathbf{W}_{L_m}\in\mathbb{C}^{L_m\times L_m}$ and $\mathbf{W}_{N}^\mathsf{H}\in\mathbb{C}^{N\times N}$ are the unitary \ac{dft} and \ac{idft} matrices, respectively. $\mathbf{P}_{L_m}\in\mathbb{N}^{L_m\times L_m}$ is the \ac{fft}-shift matrix and $\mathbf{M}_{m,r}\in\mathbb{C}^{N\times L_m}$ maps $L_m$ consecutive frequency-domain bins of the input signal to $L_m$ consecutive frequency-domain bins of the output signal as well as the rotates the phases for maintaining the phase continuity. Finally, the frequency-domain window is determined by diagonal $ \mathbf{D}_{m,r}$. For \ac{ola} scheme, the analysis and synthesis time-domain windows are given as
\ifonecolumn
\begin{subequations}
  \begin{equation}
  \mathbf{a}_{m,r} =
  \begin{bmatrix}
    \mathbf{0}_{1\times L_{\text{L},m,r}} &
    \mathbf{1}_{1\times L_{\text{S},m,r}} &
    \mathbf{0}_{1\times L_{\text{T},m,r}}
  \end{bmatrix}^\transpose
  \quad\text{and}\quad
  \mathbf{s}_{r} =
  \begin{bmatrix}
    \mathbf{1}_{N\times 1} \\
  \end{bmatrix},
\end{equation}
\else
\begin{subequations}
  \begin{equation}
  \mathbf{a}_{m,r} =
  \begin{bmatrix}
    \mathbf{0}_{L_{\text{L},m,r}\times 1} \\
    \mathbf{1}_{L_{\text{S},m,r}\times 1} \\
    \mathbf{0}_{L_{\text{T},m,r}\times 1}
  \end{bmatrix}
  \quad\text{and}\quad
  \mathbf{s}_{r} =
  \begin{bmatrix}
    \mathbf{1}_{N\times 1} \\
  \end{bmatrix},
\end{equation}
\fi
respectively, whereas for \ac{ols} scheme these windows are given by
\ifonecolumn
\begin{equation}
  \mathbf{a}_{m,r} =
  \begin{bmatrix}
    \mathbf{1}_{L_m\times 1} \\
  \end{bmatrix}
  \quad\text{and}\quad
  \mathbf{s}_{r} =
  \begin{bmatrix}
    \mathbf{0}_{1\times N_{\text{L},r}} &
    \mathbf{1}_{1\times N_{\text{S},r}} &
    \mathbf{0}_{1\times N_{\text{T},r}}
  \end{bmatrix}^\transpose,
\end{equation}
\end{subequations}
\else
\begin{equation}
  \mathbf{a}_{m,r} =
  \begin{bmatrix}
    \mathbf{1}_{L_m\times 1} \\
  \end{bmatrix}
  \quad\text{and}\quad
  \mathbf{s}_{r} =
  \begin{bmatrix}
    \mathbf{0}_{N_{\text{L},r}\times 1} \\
    \mathbf{1}_{N_{\text{S},r}\times 1} \\
    \mathbf{0}_{N_{\text{T},r}\times 1}
  \end{bmatrix},
\end{equation}
\end{subequations}
\fi
respectively.
  
The \ac{cp-ofdm} waveforms can be generated, e.g., with lowest power-of-two transform length larger than $L_{\text{act},m,n}$, i.e.,
\begin{equation}
  \label{eq:minNact}
  L_{\text{OFDM},m,n}=2^{\lceil \log_2(L_{\text{act},m,n})\rceil}
\end{equation}
and the \ac{fc} processing interpolates the \ac{cp-ofdm} waveform at the desired output rate. However, for continuous \ac{fc}-processing alternatives (as opposite to \cite{J:Yli-Kaakinen:TWC2021}), the sampling-rate conversion factor has to be selected such that the \ac{cp} length on the low-rate side is still an integer, i.e., the shortest \ac{ofdm} transform length is $L_\text{OFDM,min}=\num{128}$ corresponding to \ac{cp} length of $L_\text{CP,min}=\num{9}$ samples. Furthermore, the non-overlapping part length $L_{\text{S},m,n}$ has also to be integer, restricting the $L_m$ to be larger than equal to \num{256}.

\subsection{FC Filtered-OFDM RX Processing}
\ac{fc-f-ofdm} waveform can be received transparently, e.g., with \emph{i)} basic \ac{cp}-\ac{ofdm} receiver, \emph{ii)} by first filtering the received waveform, either using the \ac{fc}-based \ac{afb} or conventional time-domain filter, or \emph{iii)} by using \ac{wola} processing in connection with \ac{ofdm} processing.

The \ac{fc}-based \ac{afb} processing can be described as 
\begin{equation}
  \widetilde{\mathbf{y}}_m = \mathbf{G}_m \widetilde{\mathbf{z}},
\end{equation}
where the analysis processing matrix is $\mathbf{G}_m=\mathbf{F}_m^\htranspose$ and $\widetilde{\mathbf{z}}$ is the received \ac{fc-f-ofdm} waveform. Analogous to \ac{sfb} case, the decimation factor $D_m$ provided by the analysis processing is the ratio of long forward transform and short inverse transform sizes. 

Finally, the filtered and possibly decimated subband signals $\widetilde{\mathbf{y}}_m$ for $m=0,1,\dots,M-1$ are demodulated by using the conventional \ac{cp}-\ac{ofdm} \ac{rx} processing as expressed by 
\begin{subequations}
  \begin{equation}
    \widetilde{\mathbf{x}}_{m} =
    \mathbf{Q}_m
    \widetilde{\mathbf{y}}_{m},
  \end{equation}
  where
  \begin{equation}
    \mathbf{Q}_m = 
    \diag*{
      \mathbf{Q}_{m,0},\mathbf{Q}_{m,1},    
      \dots,\mathbf{Q}_{m,B_{m}-1}
    }
  \end{equation}
  with  
  \begin{equation}
    \mathbf{Q}_{m,n}=
    \widehat{\mathbf{W}}_{m,n}
    \mathbf{R}^{(\tau_{m,n})}_{N_{\text{CP},m,n}} .
  \end{equation}
  Here, $\widehat{\mathbf{W}}_{m,n}$ is the pruned unitary \ac{dft} matrix as given by \eqref{eq:prunedDFT} and $\mathbf{R}_{m,n}^{(\tau_{m,n})}\in\mathbb{Z}^{L_{\text{OFDM},m,n}\times(L_{\text{OFDM},m,n}+L_{\text{CP},m,n})}$ is the following \ac{cp} removal matrix
  \begin{align}
    \label{eq:cpRemove}
    \mathbf{R}^{(\tau_{m,n})}_{m,n} & =
    \mathbf{C}_{L_{\text{OFDM},m,n}}^{(-\tau_{m,n})}
    \begin{bmatrix}   
        \mathbf{0}_{(L_{\text{CP},m,n}-\tau_{m,n})\times L_{\text{OFDM},m,n}} \\
        \mathbf{I}_{L_{\text{OFDM},m,n}} \\
        \mathbf{0}_{\tau_{m,n}\times L_{\text{OFDM},m,n}} 
      \end{bmatrix}^\transpose,
  \end{align} 
\end{subequations}
where $\mathbf{C}^{(q)}_p$ is the circular shift matrix of size $p$ used to shift the elements of a column vector downward by $q$ elements. Here, parameter $\tau_{m,n}$ is used to control the sampling instant within the \ac{cp-ofdm} symbol as described in Section \ref{sec:wavef-requ}.

The complexity of this scheme, in terms of multiplications, is the same as for symbol-synchronized discontinuous \ac{fc} processing described in \cite{J:Yli-Kaakinen:TWC2021}, i.e., \numrange{2}{5} times the complexity of plain \ac{cp-ofdm}. The channel estimation and equalization functionalities can be realized as for the conventional \ac{cp-ofdm} waveform. 

\subsection{Frequency-Domain Window}
The frequency-domain characteristics of the \ac{fc} processing are determined by the frequency-domain window. Basically, the frequency-domain window can be adjusted at the granularity of \ac{fc} bin spacing, as given by \eqref{eq:BS}. For the proposed and the earlier approaches, the frequency-domain window consist of ones on the passband, zeros on the stopband, and two symmetric transition bands with $N_{\text{TB},m,n}$ non-trivial optimized prototype transition-band values. The same optimized transition band weights can be used for realizing all the transmission bandwidths by properly adjusting the number of one-valued weights between the transition bands. 

According to \cite[Table 5.3.3-1]{S:3GPP:TS36.104v166}, the minimum guard bands for base station channel bandwidths are determined as
\begin{equation}
  f_{\text{GB}} = 
  \frac{1}{2}\left[
  f_\text{Ch,BW}-f_{\text{SCS}}(L_\text{act,max}+1)
  \right],
\end{equation}
where $L_{\text{act,max}}=12\times N_{\text{PRB,max}}$ with $N_{\text{PRB,max}}$ being the transmission bandwidth configuration. For example, in \SI{10}{MHz} channel with \SI{15}{kHz} \ac{scs}, the maximum number of active \acp{prb} is $N_\text{PRB,max}=52$ ($L_\text{act,max}=624$) and the resulting guard band is $f_\text{GB}=\SI{312.5}{kHz}$. Assuming \ac{fc} bin spacing of $f_\text{BS}=\SI{15}{kHz}$, i.e., $N=N_\text{OFDM}=1024$, the guard band corresponds to the $f_\text{GB}/f_\text{BS}=\num{20.83}$ frequency-domain bins. Now, the frequency-domain window can be determined such that \num{624} frequency-domain window values corresponding to active subcarriers are equal to one, $\lfloor 20.83 \rfloor=20$ window values on both sides of the active subcarriers can be optimized for achieving the desired spectral characteristics while the remaining frequency-domain window values are equal to zero. Same frequency-domain window can be used for filtering the higher \acp{scs} as well, since for given channel bandwidth the guard band increases as the \ac{scs} increases. 

Let us denote the desired center frequency of each \ac{ofdm} symbol by $f^\text{(center)}_{m,n}$. The lower and higher passband (PB) edge frequencies of each symbol can then be expressed as
\ifonecolumn
  \begin{equation}    
    f_{\text{PB},m,n}^\text{(low)} =
      f^\text{(center)}_{m,n} - f_\text{s}
    \frac{L_{\text{act},m,n}/2}{L_{\text{OFDM},m,n}} 
    \qquad\text{and}\qquad
    f_{\text{PB},m,n}^\text{(high)} =
      f^\text{(center)}_{m,n} + f_\text{s}
      \frac{L_{\text{act},m,n}/2-1}{L_{\text{OFDM},m,n}},
  \end{equation}
\else
\begin{subequations}
  \begin{align}    
    f_{\text{PB},m,n}^\text{(low)} &=
      f^\text{(center)}_{m,n} - f_\text{s}
    \frac{L_{\text{act},m,n}/2}{L_{\text{OFDM},m,n}}
    \\ 
    \intertext{and}
    f_{\text{PB},m,n}^\text{(high)} &=
      f^\text{(center)}_{m,n} + f_\text{s}
      \frac{L_{\text{act},m,n}/2-1}{L_{\text{OFDM},m,n}},
\end{align}
\end{subequations}
\fi
respectively. Let us assume for simplicity that the center frequency of each subband is fixed and the subbands are sorted based on their center frequencies such that subband $m$ for $m=0$ has the lowest center frequency, subband $m$ for $m=1$ has the next lowest center frequency, and so on. Now, the stopband (SB) edge frequencies for the subbands can be expressed in the following three cases:

\subsubsection{One subband ($M=1$)} In this case, the stopband edges are determined based on the channel edges, that is, the lower and upper stopband edge frequencies are determined as
\ifonecolumn
\begin{subequations}
\begin{equation}    
  f_{\text{SB},0,n}^\text{(low)}  = {-f_\text{Ch,BW}}/{2}
  \qquad\text{and}\qquad
  f_{\text{SB},0,n}^\text{(high)} = {f_\text{Ch,BW}}/{2},
\end{equation}
\end{subequations}
\else
\begin{subequations}
\begin{align}    
  f_{\text{SB},0,n}^\text{(low)}  &= {-f_\text{Ch,BW}}/{2}
  \intertext{and}
  f_{\text{SB},0,n}^\text{(high)} &= {f_\text{Ch,BW}}/{2},
\end{align}
\end{subequations}
\fi
respectively.

\subsubsection{Two subbands ($M=2$)} In this case,  the lower (higher) stopband edge of the first (second) subband is determined by channel bandwidth while the higher (lower) stopband edge of the first (second) subband is determined by lower (higher) passband edge of the second (first) subband, that is, the lower and higher stopband edge frequencies are determined as
\begin{subequations}
\begin{equation}
  f_{\text{SB},m,n}^\text{(low)} =
  \begin{cases}
    -f_\text{Ch,BW}/2                    
      & \text{for $m=0$} \\
    f_{\text{PB},0,n}^\text{(high)} 
      & \text{for $m=1$} 
  \end{cases}
\end{equation}
and
\begin{equation}
  f_{\text{SB},m,n}^\text{(high)} =
  \begin{cases}
    f_{\text{PB},1,n}^\text{(low)} 
      & \text{for $m=0$} \\
    f_\text{Ch,BW}/2                    
      & \text{for $m=1$}. 
  \end{cases}
\end{equation}
\end{subequations}

\subsubsection{Three or more subbands ($M\geq 3$)} 
Now, the lower and higher stopband edges are determined by 
\begin{subequations}
\begin{equation}
  f_{\text{SB},m,n}^\text{(low)} =
  \begin{cases}
    -f_\text{Ch,BW}/2                    
      & \text{for $m=0$} \\
    f_{\text{PB},m-1,n}^\text{(high)} 
      & \text{for $m=1,2,\dots,M-2$} \\    
    f_{\text{PB},M-2,n}^\text{(high)}
      & \text{for $m=M-1$}
  \end{cases}
\end{equation}
and
\begin{equation}
  f_{\text{SB},m,n}^\text{(high)} =
  \begin{cases}
    f_{\text{PB},1,n}^\text{(low)}
      & \text{for $m=0$} \\
    f_{\text{PB},m+1,n}^\text{(low)} 
      & \text{for $m=1,2,\dots,M-2$} \\    
    f_\text{Ch,BW}/2                    
      & \text{for $m=M-1$},
  \end{cases}
\end{equation}
\end{subequations}
that is, the upper (lower) stopband edge of the $m$th subband is determined by the $(m+1)$th ($(m-1$)th) subband except for the edgemost subbands, where the channel edge specifies the upper (lower) stopband edge of the last (first) subband.

The frequency-domain window for $r$th \ac{fc} block on subband $m$ is determined as 
\begin{equation}
  \label{eq:win}
  \mathbf{d}_{m,r} =
  \begin{bmatrix}
    \mathbf{0}_{k_{m,n}^\text{(low)}\times 1} \\
    \mathbf{h}_{m,n} \\
    \mathbf{1}_{(k_{m,n}^\text{(high)}-k_{m,n}^\text{(low)}+1-2N_{\text{TB},m,n})\times1} \\
    \mathbf{J}_{N_{\text{TB},m,n}}\mathbf{h}_{m,n} \\
    \mathbf{0}_{(L_m-1-k_{m,n}^\text{(high)})\times 1}
  \end{bmatrix},
\end{equation}
where $n\in\{0,1,\dots,B_{\text{OFDM},m}-1\}$ is now the index of the symbol processed by the $r$th \ac{fc} block. Here, $\mathbf{h}_{m,n}\in\mathbb{R}^{N_{\text{TB},m,n}\times1}$ is the transition-band weight vector and $\mathbf{J}_{N_{\text{TB},m,n}}$ is the reverse identity matrix of size $N_{\text{TB},m,n}$ essentially reversing the order of transition-band weight vector. The lower and higher stopband indices of each subband in \eqref{eq:win} are determined as
\begin{subequations}
  \begin{equation}
    k_{m,n}^\text{(low)} = \max\left(
      \left\lceil
        \left(f^\text{(low)}_{\text{SB},m,n}-f^\text{(center)}_{m,n}\right)
        N/f_\text{s}\right\rceil
    + {L_m}/{2}, 0\right)
  \end{equation}
  and
  \begin{equation}
    k_{m,n}^\text{(high)} = \min\left(
    \left\lfloor
        \left( f^\text{(high)}_{\text{SB},m,n}-f^\text{(center)}_{m,n}\right)
      N/f_\text{s}\right\rfloor
    + {L_m}/{2}, L_m-1\right),
  \end{equation}
\end{subequations}
respectively. 

\begin{figure}[t!]             
  \centering   
  \ifonecolumn 
    \includegraphics[
    trim=20 10 50 20,clip,width=0.65\columnwidth]{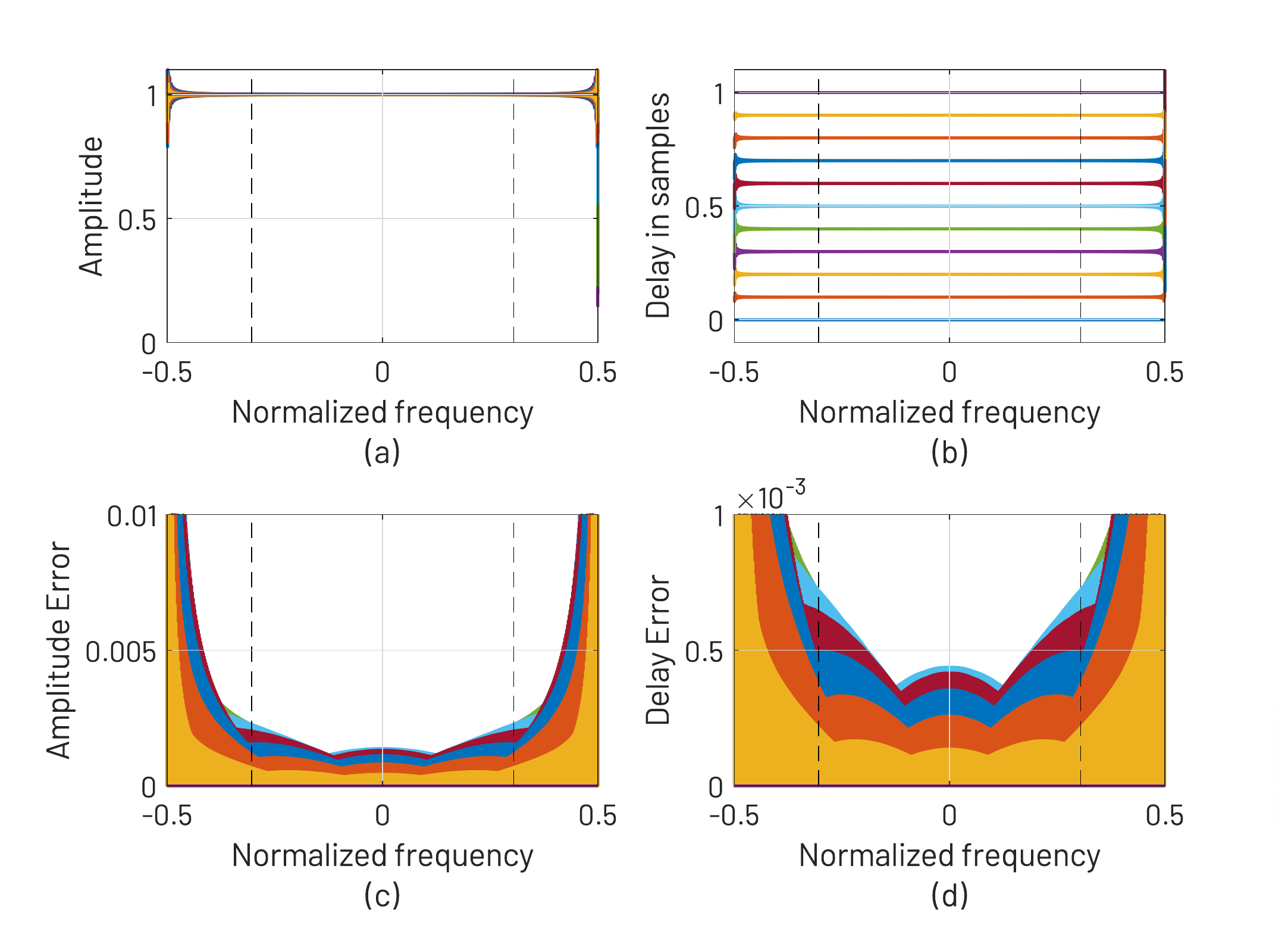}
  \else
    \includegraphics[
    trim=20 10 50 20,clip,width=0.995\columnwidth]{Figs/FloatingPointFFT_L1024.pdf}
  \fi
  \caption{Typical (a) amplitude and (b) delay responses for the \ac{fc}-based fractional delay filter for $\Phi_{\text{FD},m}\in\{0.0,0.1,\dots,1.0\}$ as well as the corresponding (c) amplitude and (d) delay errors. The frequency range between the dashed lines corresponds to the transmission bandwidth.} \label{fig:FC_FD}    
\end{figure}

For channelization purposes, the frequency-domain window values are real. \ac{fc} processing can also be used for shifting the output of each subband by fractional delay if needed.
In this case, an additional phase term as given by
\begin{equation}
  \left[\mathbf{d}_{m,r}\right]_{q} = \exp\left(-2\iu\pi(q-L_m/2)\Phi_{\text{FD},m}/L_m\right)
\end{equation}
for $q=0,1,\dots,L_m-1$ is included in the coefficients. Here, $\Phi_{\text{FD},m}\in[0,1]$ is the desired fractional delay value on subband $m$. The characteristic responses for the \ac{fc}-based fractional delay filter are illustrated in Fig.~\ref{fig:FC_FD}. 
  
\section{Waveform Requirements}%
\label{sec:wavef-requ}  
The performance of the \ac{fc-f-ofdm} waveforms are evaluated with respect to requirements defined for the \ac{5g-nr} waveform in \ac{3gpp} specification for base stations \cite{S:3GPP:TS38.104v164}. The quality of the transmitted waveform is specified by the \ac{evm} requirements, defining the maximum allowable deviation of the transmitted symbols with respect to ideal ones. The \ac{oobe} requirements, on the other hand, give the requirements for the tolerable spectral emissions of \ac{tx} waveform. In addition to these two key metrics, there are other measures, e.g.,  \ac{evm} equalizer flatness requirements, \acp{ibe}, \ac{obw}, among others, however, these measures are beyond the scope of this paper.

\subsection{Error Vector Magnitude}
\begin{figure}[t!]               
  \centering   
  \ifonecolumn 
    \includegraphics[width=0.65\columnwidth]{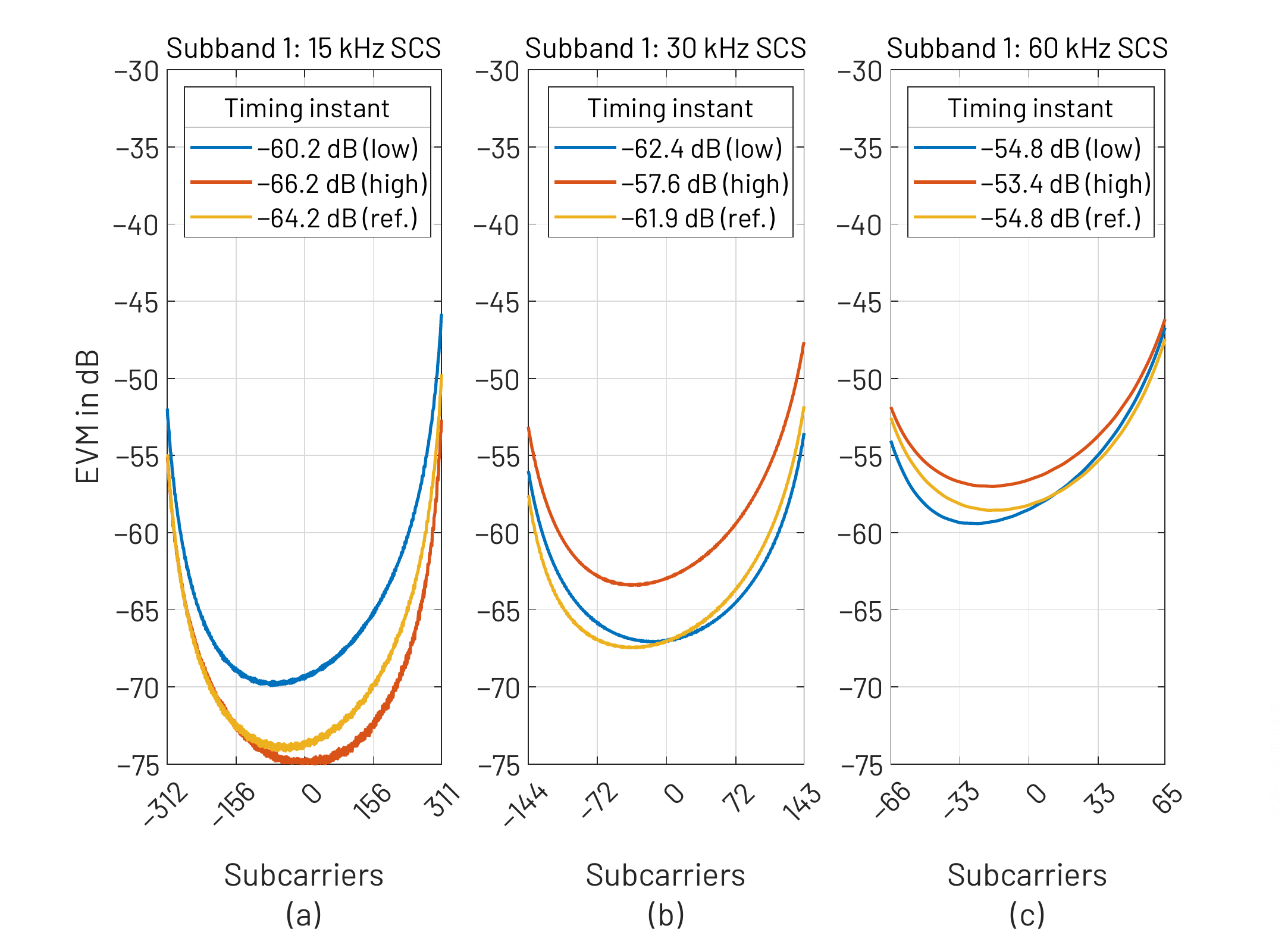}
  \else  
    \includegraphics[
    trim=30 0 50 0,clip,width=\columnwidth]{Figs/EVM_M2_10MHz.pdf} 
  \fi   
  \caption{\acsp{evm} for the example mixed-numerology scenario in Fig.~\ref{fig:MixedBlock}. (a) \SI{15}{kHz} \ac{scs} with \num{624} active SCs, (b) \SI{30}{kHz} \ac{scs} with \num{288} active SCs, and (c) \SI{60}{kHz} \ac{scs} with \num{132} active SCs. \acp{evm} are evaluated with the reference timing (ref.) as well as $N_\text{EVM}/2$ samples before (low) and after (high) the reference timing. \ac{5g-nr} \ac{evm} window values from \cite[Tables B.5.2-1, B.5.2-2, and B.5.2-3]{S:3GPP:TS38.104v164} are used. The average \acp{evm} over the active subcarriers are shown at the legend.}    
  \label{fig:EVMexample} 
\end{figure}

The quality of the \ac{tx} processing in \ac{5gnr} is measured by evaluating the \ac{mse} between the transmitted and ideal symbols. For the proposed approach, $\Upsilon_{m}$ symbol sets with different numerology are allowed on subband $m$ and, therefore, the \ac{mse} is evaluated separately for each numerology as 
\begin{equation}
  \label{eq:mse} 
  \boldsymbol{\epsilon}_{\text{MSE},m,\upsilon}=
  \frac{1}{\lvert\mathcal{S}_ {m,\upsilon}\rvert}
  \sum_{n \in \mathcal{S}_{m,\upsilon}}
  \lvert 
  \widetilde{\mathbf{x}}_{m,n}-\mathbf{x}_{m,n}
  \rvert^2
\end{equation} 
for $\upsilon=0,1,\dots,\Upsilon_m-1$ as illustrated in Fig.~\ref{fig:EVMexample}. The corresponding \acf{evm} in percents is expressed as 
\begin{equation} 
  \boldsymbol{\epsilon}_{\text{EVM},m,\upsilon} =
  100\sqrt{\boldsymbol{\epsilon}_{\text{MSE},m,\upsilon}}.
\end{equation}
In this contribution, the \ac{evm} is expressed in decibels as
\begin{equation}
  \boldsymbol{\epsilon}_{\text{EVM},m,\upsilon}
  =10\log_{10}(\boldsymbol{\epsilon}_{\text{MSE},m,\upsilon}).
\end{equation}
Here, \ac{mse} and \ac{evm} are measured after executing \ac{zfe}, as defined in \cite[Annex B]{S:3GPP:TS38.104v164}.

The average \ac{mse} is defined as the arithmetic mean of the \ac{mse} values on active subcarriers, as given by
\begin{equation} 
  \label{eq:MSEavg}
  \bar{\epsilon}_{\text{MSE},m,\upsilon}=
    \frac{1}{L_{\text{act},\upsilon}}
    \mathbf{1}_{1\times L_{\text{act},v}}
    \boldsymbol{\epsilon}_{\text{MSE},m,\upsilon}
\end{equation}
Here, $L_{\text{act},\upsilon}$ is the number of active subcarriers for the symbols in index set $\mathcal{S}_{m,\upsilon}$, that is, $L_{\text{act},m,n}$ for $n\in\mathcal{S}_{m,\upsilon}$. The corresponding average \ac{evm} is denoted by $\bar{\epsilon}_{\text{EVM},m,\upsilon}$.
 
In general, \ac{evm} can be measured at $N_{\text{CP},m,n}$ timing instances by modifying the \ac{cp} removal matrix as expressed by \eqref{eq:cpRemove}. In this case, the timing adjustment has to be compensated by circularly shifting the \ac{ofdm} symbols before taking the \ac{fft}. According to \cite{S:3GPP:TS38.104v164}, the timing instant in the middle of the \ac{cp} is selected as a reference point and the \ac{evm} performance before and after the reference point is measured in order to characterize the \ac{evm} performance degradation with respect to timing errors. 

\begin{figure}[t!]             
  \centering    
  \ifonecolumn 
    \includegraphics[width=0.6\columnwidth]{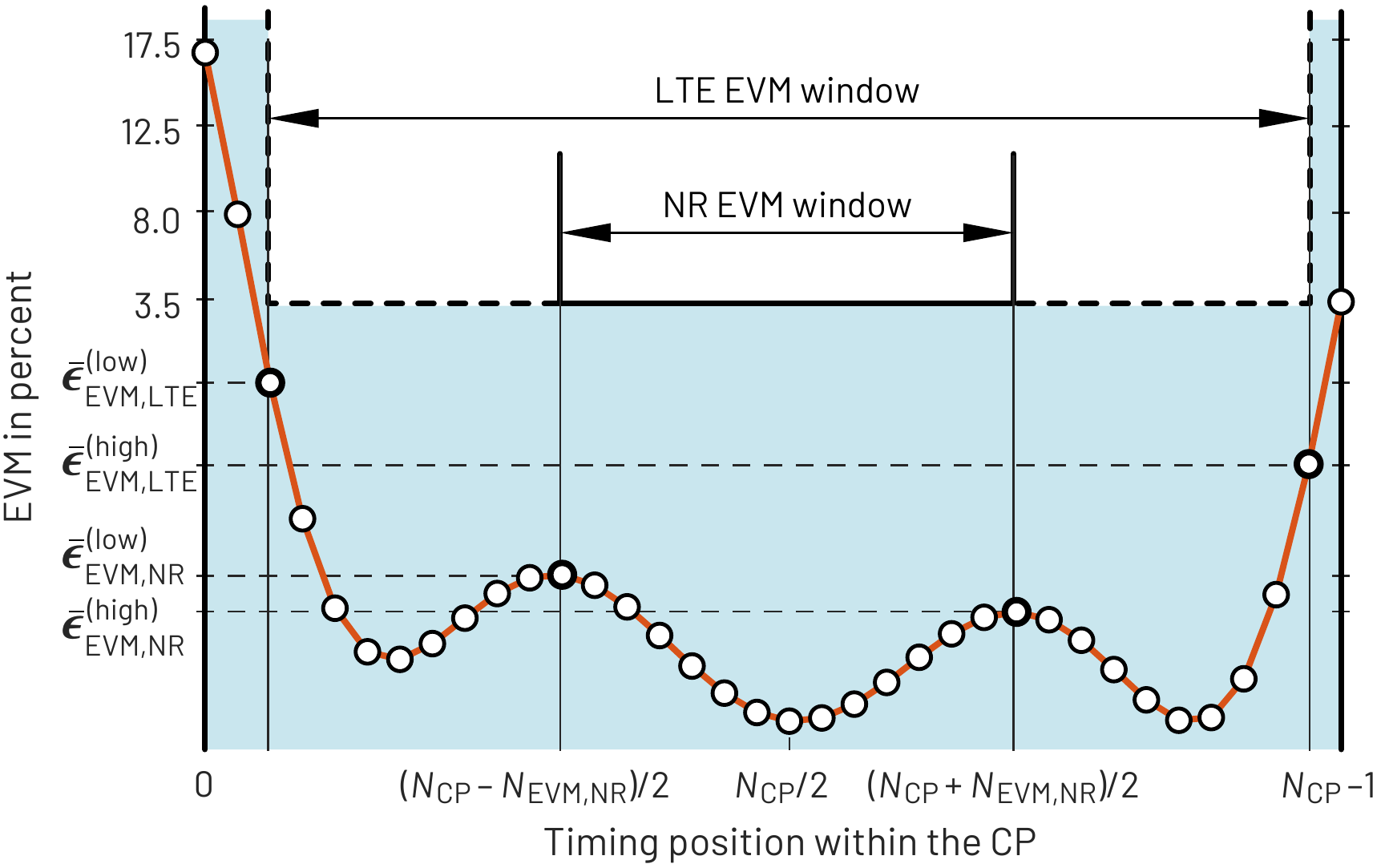}
  \else
    \includegraphics[width=\figwidth]{Figs/EVMmask2021Nov12.pdf}
  \fi
  \caption{\acs{evm} is measured at both sides of the ideal timing instance where the measurement positions are defined by the \acs{evm} window length. The \ac{lte} \ac{evm} window is considerably longer than the \ac{5g-nr} \ac{evm} window.}   
  \label{fig:EVMwin} 
\end{figure}   

Fig.~\ref{fig:EVMwin} illustrates the \ac{evm} evaluation for \ac{lte} and \ac{5g-nr} waveforms. In the case of \ac{5g-nr} waveform, the \ac{evm} is measured $N_{\text{EVM,NR}}/2$ samples before and after the reference point, where $N_{\text{EVM,NR}}$ is the \ac{evm} window length, and the corresponding \ac{evm} values are denoted as $\bar{\epsilon}_\text{EVM,NR}^\text{(low)}$ and $\bar{\epsilon}_\text{EVM,NR}^\text{(high)}$, respectively. The requirements for the \ac{evm} can be interpreted in the context of the \ac{evm} requirements of \ac{5gnr}, stated as \{\SI{17.5}{\%}, \SI{12.5}{\%}, \SI{8.0}{\%}, \SI{3.5}{\%}\} or \{\SI{-15}{dB}, \SI{-18}{dB}, \SI{-22}{dB}, \SI{-29}{dB}\} for {\acs{qpsk}, 16-\acs{qam}, 64-\acs{qam}, 256-\acs{qam}}, respectively \cite[Table 6.5.2.2-1]{S:3GPP:TS38.104v164}.

For \ac{lte},  $\bar{\epsilon}_\text{EVM,LTE}^\text{(low)}$ and $\bar{\epsilon}_\text{EVM,LTE}^\text{(high)}$ are evaluated in a same manner, however, while the \ac{5g-nr} \ac{evm} window lengths are \SIrange{40}{60}{\%} of the \ac{cp} length \cite[Tables B.5.2-1--B.5.2-3 for FR1]{S:3GPP:TS38.104v164}, the corresponding \ac{lte} \ac{evm} windows are considerably longer, that is, \SIrange{55.6}{94.4}{\%} \cite[Table E.5.1-1]{S:3GPP:TS36.104v166} implying relaxed time-synchronization requirements although more stringent requirements for waveform purity.

\subsection{Unwanted emissions}
In the base station case, \acfp{oobe} are unwanted emissions immediately outside the channel bandwidth. The \ac{oobe} requirements for the base station transmitter are specified both in terms of \ac{obue} and \ac{aclr}. \ac{obue} define all unwanted emissions in each supported downlink operating band as well as the frequency ranges $\Delta{f_\text{OBUE}}$ above and $\Delta{f_\text{OBUE}}$ below each band. In \ac{5gnr}, $\Delta{f_\text{OBUE}=\SI{10}{MHz}}$ and $\Delta{f_\text{OBUE}=\SI{40}{MHz}}$ in \ac{fr1} and \ac{fr2}, respectively \cite[Table 6.6.1-1]{S:3GPP:TS38.104v164}. \ac{aclr} is the ratio of the filtered mean power centred on the assigned channel frequency to the filtered mean power centred on an adjacent channel frequency \cite{S:3GPP:TS38.104v164}. 
 
\begin{figure}[t!]                 
  \centering   
  \ifonecolumn 
    \includegraphics[width=0.6\columnwidth]{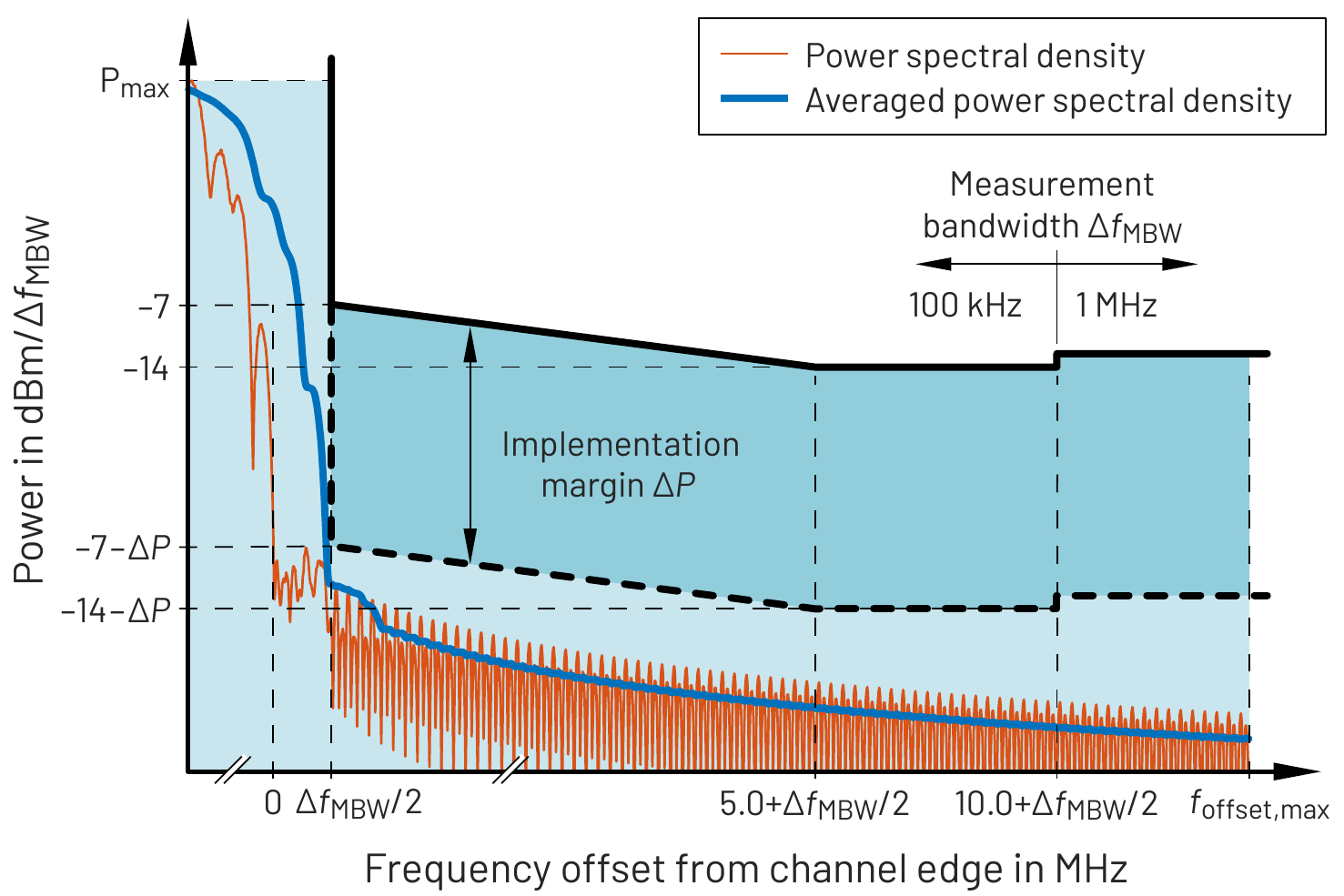}
  \else  
    \includegraphics[
    trim=0 0 0 0,clip,width=0.94\columnwidth]{Figs/5GNR-SEMoffset2021Aug14.pdf} 
  \fi  
  \caption{Example \acp{psd} and typical \ac{obue} requirements (following \cite[Table 6.6.4.2.1-2]{S:3GPP:TS38.104v164}). Here, the maximum \ac{psd} level $P_\text{max}$ is determined based on the maximum output power of the base station.}    
  \label{fig:OBUElimits}  
\end{figure}

Fig.~\ref{fig:OBUElimits} illustrates typical \ac{obue} requirements adopting the limits defined in \cite[Table 6.6.4.2.1-2]{S:3GPP:TS38.104v164}. Thin (red) response shows the non-averaged \ac{psd} estimate and thick (blue) solid response shows the averaged \ac{psd} estimate with \ac{mbw} of $\Delta{f_\text{MBW}}$. Now, the requirements are stated such that for frequency offset of $f_\text{offset}=\Delta{f_\text{MBW}}/2$ from the channel edge, the maximum allowed power for the averaged \ac{psd} estimate is \SI{-7}{dBm}. This requirement increases linearly to \SI{-14}{dBm} for frequency offset of $f_\text{offset}=\SI{5}{MHz}+\Delta{f_\text{MBW}}/2$ and maintains a constant value until $f_\text{offset}=\SI{10}{MHz}+\Delta{f_\text{MBW}}/2$. The maximum carrier output power of the (wide-area) base station is vendor specific, e.g., $P_\text{max}=\SI{33}{dBm}$ in \SI{10}{MHz} channel, translating into \SI{40}{dB} attenuation requirement at the channel edge with respect to in-band level for the waveform generation. In addition to these requirements, some additional margin is needed to cope with performance degradation due to implementation non-idealities, i.e., finite-precision arithmetic, \ac{pa} non-linearity, etc.
Similarly, on the \ac{rx} side, certain level of frequency selectivity is needed in order to limit the \ac{ini} as well as to provide sufficient rejection from \acs{rf} blockers and other interferences.
 
The \ac{psd} estimate (or \emph{the sample spectrum}) of the transmitted waveform can be evaluated by taking the \ac{dft} of the time-domain waveform and then squaring the absolute value of the resulting frequency-domain response as given by \cite{C:Djuric99}
\begin{equation}
  \tilde{\mathbf{s}}_\textbf{z} =
  \frac{1}{N_{\text{PSD}}}
  \left\lvert 
    \mathbf{W}_{N_{\text{PSD}}}
    \begin{bmatrix}
      \mathbf{z} \\
      \mathbf{0}_{(N_\text{PSD}-N_\text{samp})\times 1}
    \end{bmatrix}
  \right\rvert^2.  
\end{equation}
Here, the signal is first zero padded to desired (e.g., power-of-two) length $N_{\text{PSD}}$ and $\mathbf{W}_{N_{\text{PSD}}}$ is the \ac{dft} matrix of size $N_{\text{PSD}}$. The \ac{rbw} of the non-averaged \ac{psd} estimate is
\begin{equation}
  \Delta_\text{RBW} = f_\text{s}/N_{\text{PSD}}.
\end{equation}
The \ac{psd} estimate for a given \ac{mbw} $\Delta_\text{MBW}$ can obtained by averaging neighboring $N_\text{avg}=\Delta_\text{MBW}/\Delta_\text{RBW}$ spectral estimates \cite{B:Bendat10}. Moving-average filter can be conveniently realized using frequency-domain element-wise multiplication as
\begin{equation} 
  \bar{\mathbf{s}}_\textbf{z} =   \mathbf{W}^\mathsf{H}_{N_{\text{PSD}}}\left(
   [\mathbf{W}_{N_{\text{PSD}}}\mathbf{r}]
   \odot
   [\mathbf{W}_{N_{\text{PSD}}}\tilde{\mathbf{s}}_{\textbf{z}}]
  \right),
\end{equation}
where $\odot$ denotes the element-wise multiplication and the rectangular 
moving-average filter kernel is given by
\begin{equation}
  \mathbf{r} = 
  \begin{bmatrix}
    \mathbf{1}_{\lceil N_\text{avg}/2\rceil \times 1}   \\
    \mathbf{0}_{(N_\text{PSD}-N_\text{avg}) \times 1}   \\
    \mathbf{1}_{\lfloor N_\text{avg}/2\rfloor \times 1}    
  \end{bmatrix}.
\end{equation}

Adopting the specifications in \cite{S:3GPP:TS38.104v164,S:3GPP:TS36.104v166}, the \ac{mbw} of $\Delta_{\text{MBW}}=\SI{100}{kHz}$ is commonly used for \ac{5g-nr} bands below \SI{1}{GHz} whereas for \ac{5g-nr} bands above \SI{1}{GHz}, $\Delta_{\text{MBW}}=\SI{1}{MHz}$ is also used for large frequency offsets from measurement filter centre frequency. 

\section{Numerical Examples}
\label{sec:examples}
The performance and the flexibility of the proposed processing is demonstrated in terms of four examples. In all examples, \ac{fc}-based filtering is also used on the \ac{rx} side prior to \ac{ofdm} demodulation if not  stated otherwise.

\subsection{Channelization of \acfp{bwp}}
In this example, we demonstrate the division of a channel into non-contiguous \acp{bwp} with mixed-numerology. We consider \SI{50}{MHz} channel with four ($M=4$) \acp{bwp} such that transmission bandwidth is divided into \SI{5}{MHz}, \SI{10}{MHz}, \SI{20}{MHz}, and \SI{15}{MHz} \acp{bwp} with \SI{30}{kHz}, \SI{15}{kHz}, \SI{60}{kHz}, and \SI{30}{kHz} \acp{scs}, respectively. According to \cite[Table 5.3.2-1]{S:3GPP:TS36.104v166}, the number of active subcarriers are $L_{\text{act},0,n}=132$ for $n=0,1,\dots,13$ (11 \acp{prb}), $L_{\text{act},1,n}=624$ for $n=0,1,\dots,6$ (52 \acp{prb}), $L_{\text{act},2,n}=288$ for $n=0,1,\dots,27$ (24 \acp{prb}), and $L_{\text{act},3,n}=456$ for $n=0,1,\dots,13$ (38 \acp{prb}), respectively. 

\begin{figure}[t!]                 
  \centering   
  \ifonecolumn 
    \includegraphics[width=0.65\columnwidth]{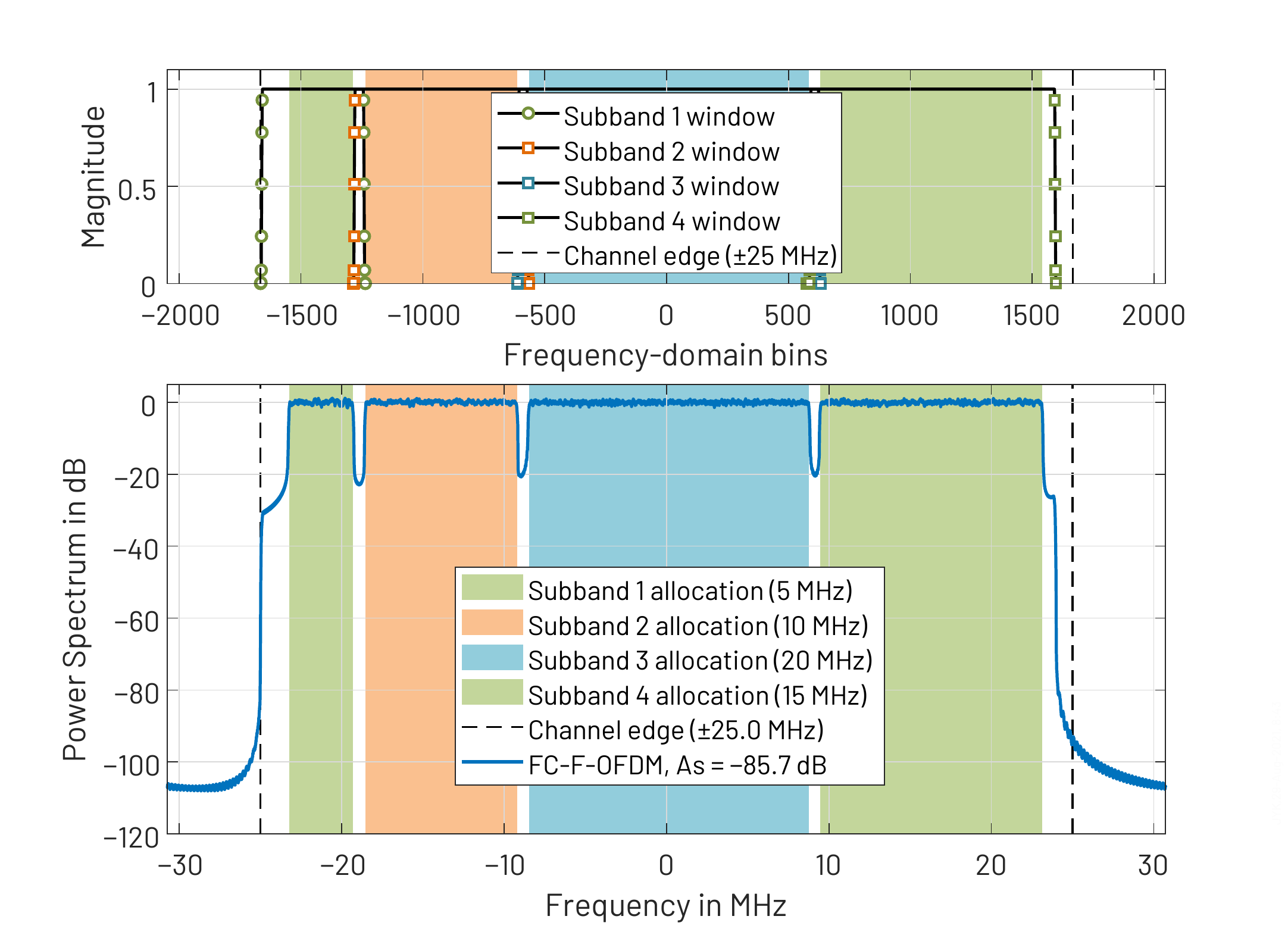}
  \else  
    \includegraphics[
    trim=30 0 45 0,clip,width=\columnwidth]{Figs/PSD_M4_50MHz.pdf} 
  \fi  
  \caption{\acs{psd} of the composite waveform in Example 1 consisting of four \acp{bwp} and the corresponding frequency-domain windows.}    
  \label{fig:Ex1:PSD}  
\end{figure}  

The \ac{ofdm} transform lengths needed to obtain sampling rate of $f_\text{s}=\SI{61.44}{MHz}$ are $2N_{\text{OFDM},0,n}=4N_{\text{OFDM},1,n}= N_{\text{OFDM},2,n}=2N_{\text{OFDM},3,n}=1024$. However, the complexity can be reduced by using interpolating \ac{fc} processing with shortest power-of-two \ac{ofdm} transform lengths for given number of active subcarriers, as given by \eqref{eq:minNact}. In this case, the \ac{ofdm} transform lengths on the low-rate side can be reduced to $L_{\text{OFDM},0,n}=2L_{\text{OFDM},1,n}= L_{\text{OFDM},2,n}=L_{\text{OFDM},3,n}=512$, by using interpolating \ac{fc} processing with $2I_0=2I_1=I_2=2I_3=2$, i.e., the \ac{fc}-processing inverse and forward transform lengths are selected as $4N=L_0=L_1=2L_2=L_3=1024$. 

In this example, we have used guard band of \SI{720}{kHz} (4~\acp{prb} with \SI{15}{kHz} \ac{scs}) between the \acp{bwp} such that the center frequencies of the \acp{bwp} are $f^\text{(center)}_{0,n}=\SI{-21240}{kHz}$, $f^\text{(center)}_{1,n}=\SI{-13860}{kHz}$, $f^\text{(center)}_{2,n}=\SI{180}{kHz}$, and $f^\text{(center)}_{3,n}=\SI{16380}{kHz}$, respectively. The sub-modulations used on \SI{5}{MHz}, \SI{10}{MHz}, \SI{20}{MHz}, and \SI{15}{MHz} \acp{bwp}  are 64-\ac{qam}, 16-\ac{qam}, \ac{qpsk}, and 16-\ac{qam}, respectively. 

\begin{figure}[t!]                 
  \centering   
  \ifonecolumn 
    \includegraphics[width=0.65\columnwidth]{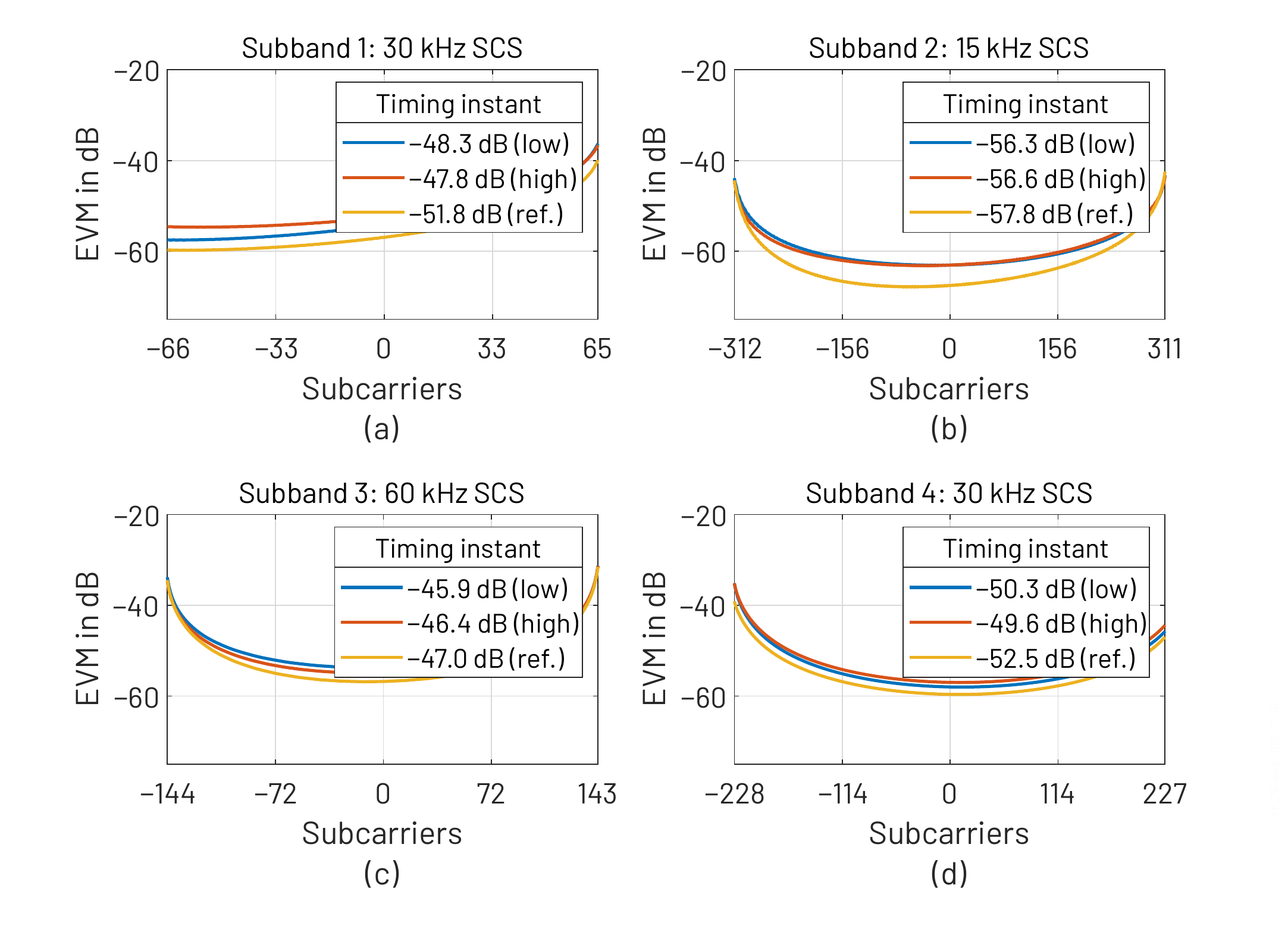}
  \else  
    \includegraphics[
    trim=30 0 45 0,clip,width=\columnwidth]{Figs/EVM_M4_50MHz.pdf} 
  \fi  
  \caption{\acs{evm} on each subband in Example A. (a) \SI{5}{MHz} carrier with \SI{30}{kHz} \ac{scs} and 132 active \acp{sc}. (b) \SI{10}{MHz} carrier with \SI{15}{kHz} \ac{scs} and 624 active \acp{sc}. (c) \SI{20}{MHz} carrier with \SI{60}{kHz} \ac{scs} and 288 active \acp{sc}. (d) \SI{15}{MHz} carrier with \SI{30}{kHz} \ac{scs} and 456 active \acp{sc}.}    
  \label{fig:Ex1:EVM} 
\end{figure}

The \ac{psd} of the resulting \ac{fc}-based filtered-\ac{ofdm} waveform is shown in Fig.~\ref{fig:Ex1:PSD} and the \acp{evm} for the \acp{bwp} are shown in Fig.~\ref{fig:Ex1:EVM}. The emission level at channel edge is $A_\text{s}=-\SI{85.7}{dB}$ and the simulated \acp{evm} values for the \acp{bwp} are $\bar{\epsilon}_{\text{EVM},0,0}=\SI{-51.8}{dB}$, $\bar{\epsilon}_{\text{EVM},1,0}=\SI{-57.8}{dB}$, $\bar{\epsilon}_{\text{EVM},2,0}=\SI{-47.0}{dB}$, and $\bar{\epsilon}_{\text{EVM},3,0}=\SI{-52.5}{dB}$.
The \ac{evm} values at \ac{evm} low and high timing instances are shown in Fig.~\ref{fig:Ex1:EVM}. As seen, the simulated \ac{evm} values are at least \SI{25}{dB} better than the requirements stated in \cite[Table 6.5.2.2-1]{S:3GPP:TS38.104v164}.

For reference, the corresponding emission level for the \ac{wola}-based \ac{tx} is $A_\text{s}=\SI{-60.7}{dB}$ while the simulated \acp{evm} values for the \acp{bwp} when \ac{wola} is used on \ac{tx} and \ac{rx} sides are $\bar{\epsilon}_{\text{EVM},0,0}=\SI{-31.9}{dB}$, $\bar{\epsilon}_{\text{EVM},1,0}=\SI{-34.0}{dB}$, $\bar{\epsilon}_{\text{EVM},2,0}=\SI{-30.8}{dB}$, and $\bar{\epsilon}_{\text{EVM},3,0}=\SI{-33.0}{dB}$. In this case, \ac{wola} extension lengths are $L_\text{ext}=L_{\text{CP},m,n}/4$.

\subsection{Wide-Band Carrier with Guard-Band IoT} 
In this example, we demonstrate the co-existence of \ac{5g-nr} wide-band carrier and \ac{4g}-based narrow-band (NB) \ac{iot}\acused{nb-iot} carriers in a same channel. Here, we consider \SI{20}{MHz} channel with \ac{nb-iot} on the guard-band of the wide-band carrier. The wide-band carrier has $L_{\text{act},0,n}=312$ active \acp{sc} with \SI{30}{kHz} \acp{scs} for $n=0,1,\dots,13$ while the NB-\ac{iot} carriers have $L_{\text{act},1,n}=L_{\text{act},2,n}=L_{\text{act},3,n}=L_{\text{act},4,n}=12$ \acp{sc} with \SI{15}{kHz} \acp{scs} for $n=0,1,\dots,6$. In this case, each pair of \ac{nb-iot} carriers is filtered as a single subband such that the total number of subbands (and frequency-domain windows) is three. The guard band between the narrow-band carriers with \SI{15}{kHz} \ac{scs} and wide-band carrier with \SI{30}{kHz} \ac{scs} is \SI{180}{kHz}. In this case, 64-\ac{qam} is used on the wide-band carrier and \ac{qpsk} on \ac{nb-iot} carriers. \ifonecolumn\else The time-frequency allocation of this example is illustrated in Fig.~\ref{fig:Ex2:TFA}. \fi The \ac{ofdm} transform lengths are now $4L_{\text{OFDM},0,n}=L_{\text{OFDM},1,n}= L_{\text{OFDM},2,n}=256$, that is, the \ac{iot} subcarriers are interpolated by eight while the \ac{fc}-processing transform lengths are $N=8L_0=L_1=L_2=256$.
    
The \ac{psd} of the resulting \ac{fc}-based filtered-\ac{ofdm} waveform is shown in Fig.~\ref{fig:Ex2:PSD} and the \acp{evm} for the subbands are shown in Fig.~\ref{fig:Ex2:EVM}. The power level at channel edge is $A_s=\SI{-78.1}{dB}$ and the simulated \ac{evm} values for the carriers are $\bar{\epsilon}_{\text{EVM},0,0}=\SI{-48.9}{dB}$, $\bar{\epsilon}_{\text{EVM},1,0}=\SI{-39.7}{dB}$, and $\bar{\epsilon}_{\text{EVM},2,0}=\SI{-44.7}{dB}$. Here, the \ac{evm} values of each pair of \ac{nb-iot} carriers are combined for simplicity. Again, the simulated \ac{evm} values are at least \SI{25}{dB} above the requirements. The corresponding \ac{evm} low and high values are shown in Fig.~\ref{fig:Ex2:EVM}. 

\ifonecolumn 
\else
  \begin{figure}[t!]                 
  \centering   
  \ifonecolumn 
    \includegraphics[width=0.65\columnwidth]{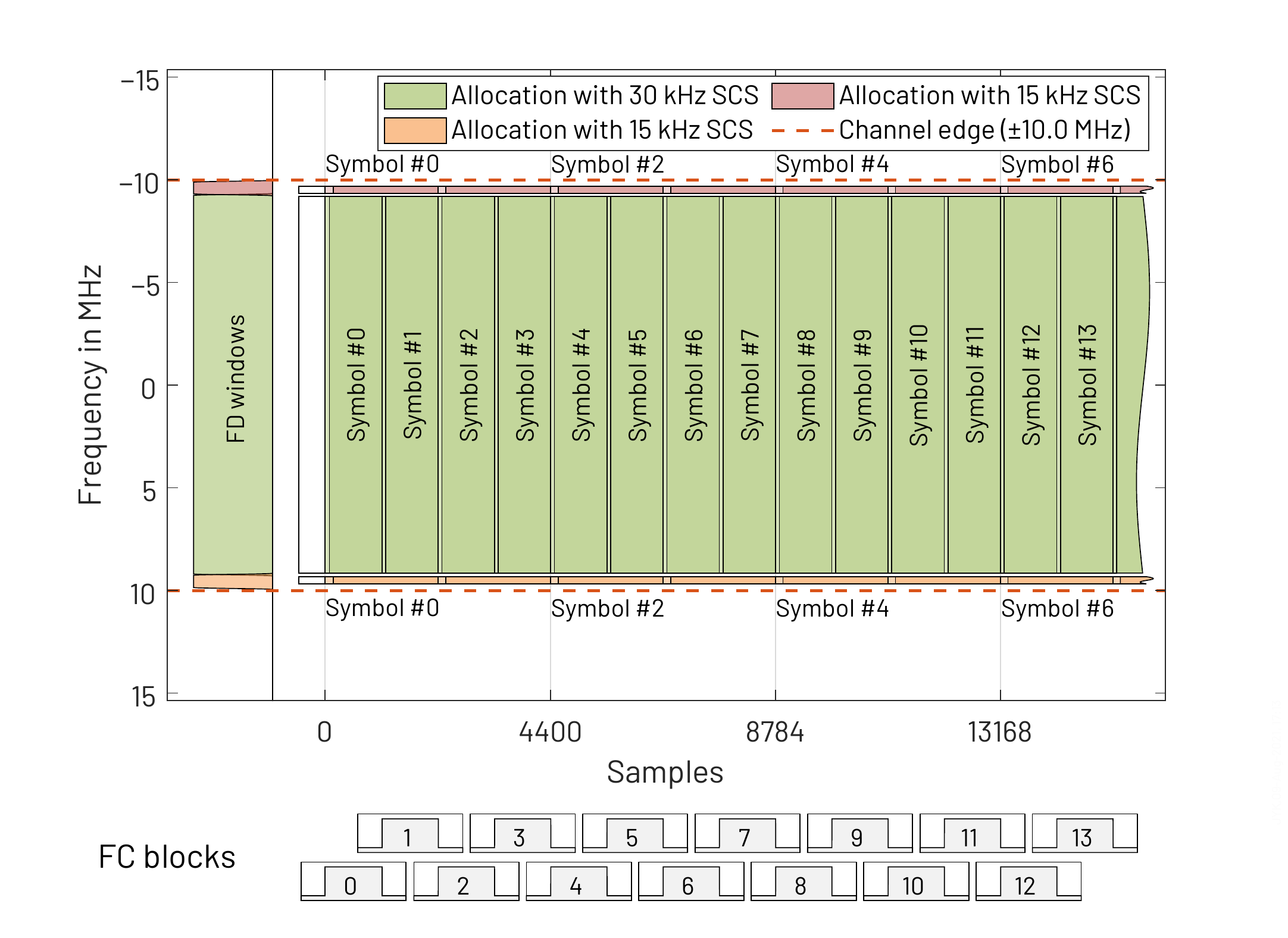}
  \else  
    \includegraphics[
    trim=30 0 50 0,clip,width=0.95\columnwidth]{Figs/TFA_M3_20MHz.pdf} 
  \fi    
  \caption{Time-frequency allocation in Example B with three subbands.}    
  \label{fig:Ex2:TFA} 
\end{figure}
end
 
\begin{figure}[t!]                 
  \centering    
  \ifonecolumn 
    \includegraphics[width=0.65\columnwidth]{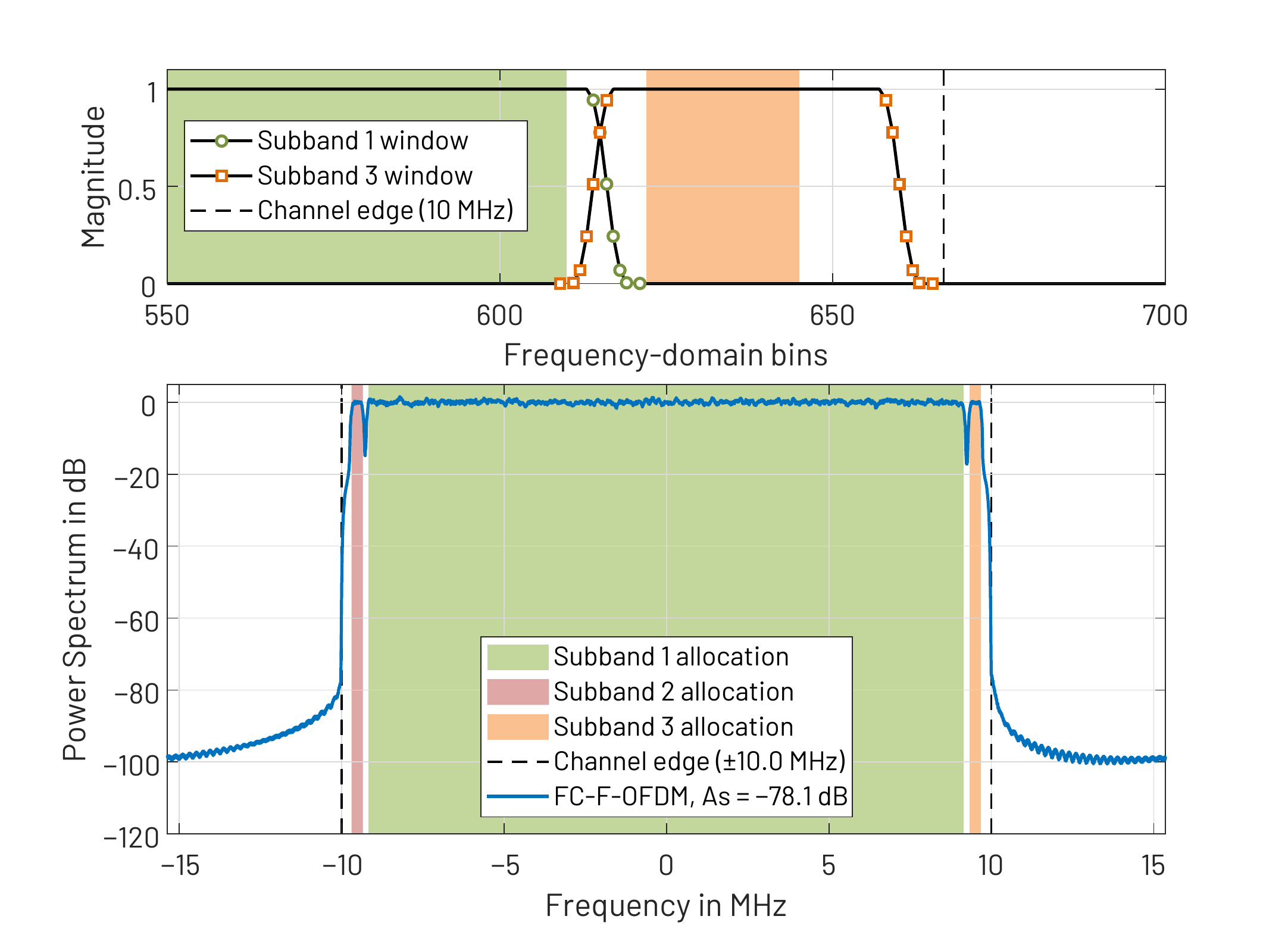}
  \else  
    \includegraphics[
    trim=30 0 45 0,clip,width=\columnwidth]{Figs/PSD_M3_20MHz.pdf} 
  \fi  
  \caption{\acs{psd} of the composite waveform and details of the frequency-domain windows in Example B.}    
  \label{fig:Ex2:PSD}  
\end{figure}  
 
\begin{figure}[t!]                 
  \centering   
  \ifonecolumn 
    \includegraphics[width=0.65\columnwidth]{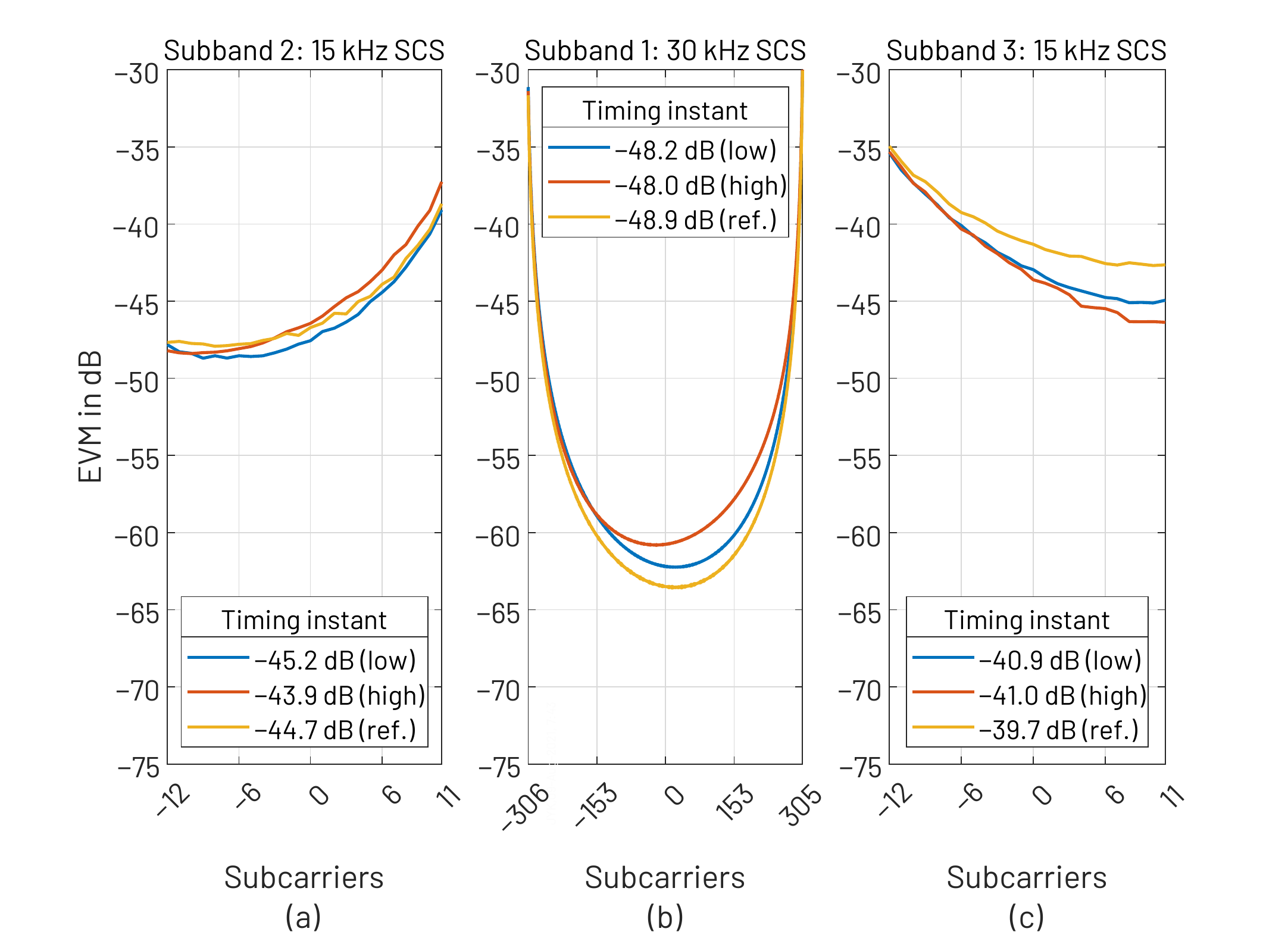}
  \else  
    \includegraphics[
    trim=30 0 45 0,clip,width=\columnwidth]{Figs/EVM_M3_20MHz.pdf} 
  \fi  
  \caption{\acs{evm} on each three subbands in Example B. (a) Leftmost subband containing two \ac{nb-iot} carriers with \SI{15}{kHz} \ac{scs}. (a) \SI{20}{MHz} carrier with \SI{30}{kHz} \ac{scs}. (c) Rightmost subband containing two \ac{nb-iot} carriers with \SI{15}{kHz} \acp{scs}.}    
  \label{fig:Ex2:EVM}  
\end{figure} 
  
\subsection{\acs{ssb}-like Mixed-Numerology Scenario}
In this example, we consider a \ac{ssb}-like scenario where a wide-band carrier with one \ac{scs} is punctured by the symbols with another \ac{scs}. In this case, four \ac{ofdm} symbols with \SI{15}{kHz} \ac{scs} ($L_{\text{act},0,n}=624$ active \acp{sc} for $n=0,1,2,3$) and six symbols with \SI{30}{kHz} \ac{scs} ($L_{\text{act},0,n}=288$ active \acp{sc} for $n=4,5,\dots,9$) are transmitted in \SI{10}{MHz} channel. Second and third symbol with \SI{15}{kHz} \ac{scs} is punctured by the eight \ac{ofdm} symbols with \SI{60}{kHz} \ac{scs} (120 active \acp{sc}) such that the $528$ innermost subcarriers of the symbols with \SI{15}{kHz} \ac{scs} are deactivated. Third and fourth \ac{ofdm} symbol with \SI{30}{kHz} \ac{scs} is punctured by one \ac{ofdm} symbol with \SI{15}{kHz} \ac{scs} (432 active \acp{sc}) such that 240 \acp{sc} of the symbols with \SI{30}{kHz} \acp{scs} are deactivated. The resulting guard-band between the allocations with different \acp{scs} is about \SI{360}{kHz}. In this example, \ac{qpsk} is used for all allocations. The time-frequency allocation of this mixed-numerology scenario is detailed in Fig.~\ref{fig:Ex3:TFA}. 
 
The simulated power level at the channel edge is \SI{76.9}{dB}. The \acp{evm} for the symbols for each numerology are shown in Fig.~\ref{fig:Ex3:EVM}. As seen, the symbols with \SI{60}{kHz} \acp{scs} have the worst \ac{evm} since, for these subcarriers, the guard-band relative to \ac{scs} is the smallest. The peaks seen in the \ac{evm} responses of Fig.~\ref{fig:Ex3:EVM}(a) are due to the time-domain transients resulting from filtering the symbols with \SI{60}{kHz} \ac{scs}. Similarly, the peaks in Fig.~\ref{fig:Ex3:EVM}(b) are the transients of allocation with 432 \acp{sc}, i.e., the improved frequency-domain localization increases the dispersion in time domain, however, even with this \ac{isi}, the \ac{evm} levels are still within the requirements.  

\begin{figure}[t!]                 
  \centering   
  \ifonecolumn 
    \includegraphics[width=0.65\columnwidth]{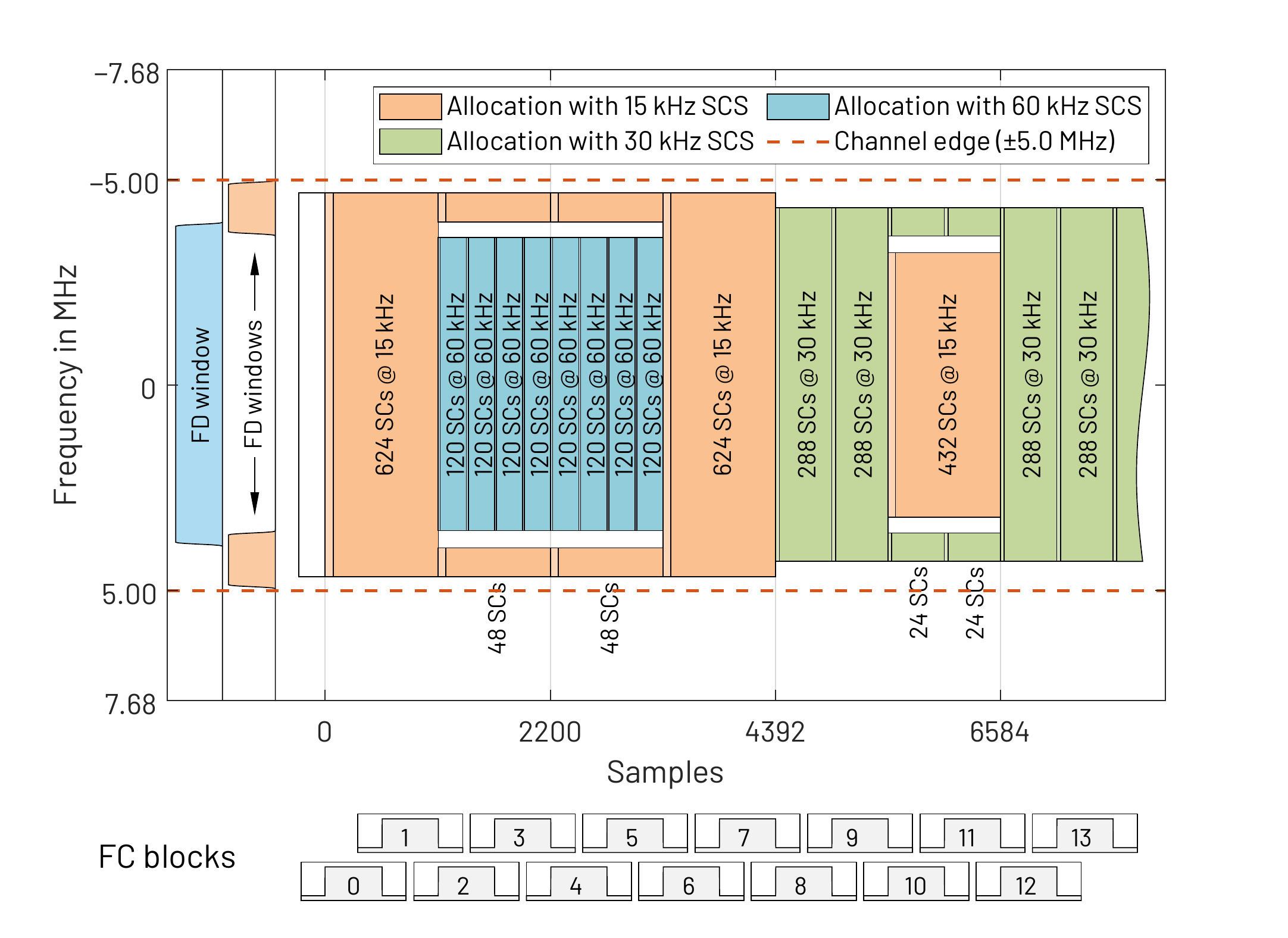}
  \else  
    \includegraphics[
    trim=20 0 50 0,clip,width=0.95\columnwidth]{Figs/TFA_M6_10MHz_2.pdf} 
  \fi
  \caption{Time-frequency allocation in Example C with mixed numerology. The frequency-domain windows during the $r$th \ac{fc} block for $r=2,3,4,5$ are illustrated on the left hand side of the figure.}     
  \label{fig:Ex3:TFA} 
\end{figure}  
  
\begin{figure}[t!]                 
  \centering   
  \ifonecolumn 
    \includegraphics[width=0.65\columnwidth]{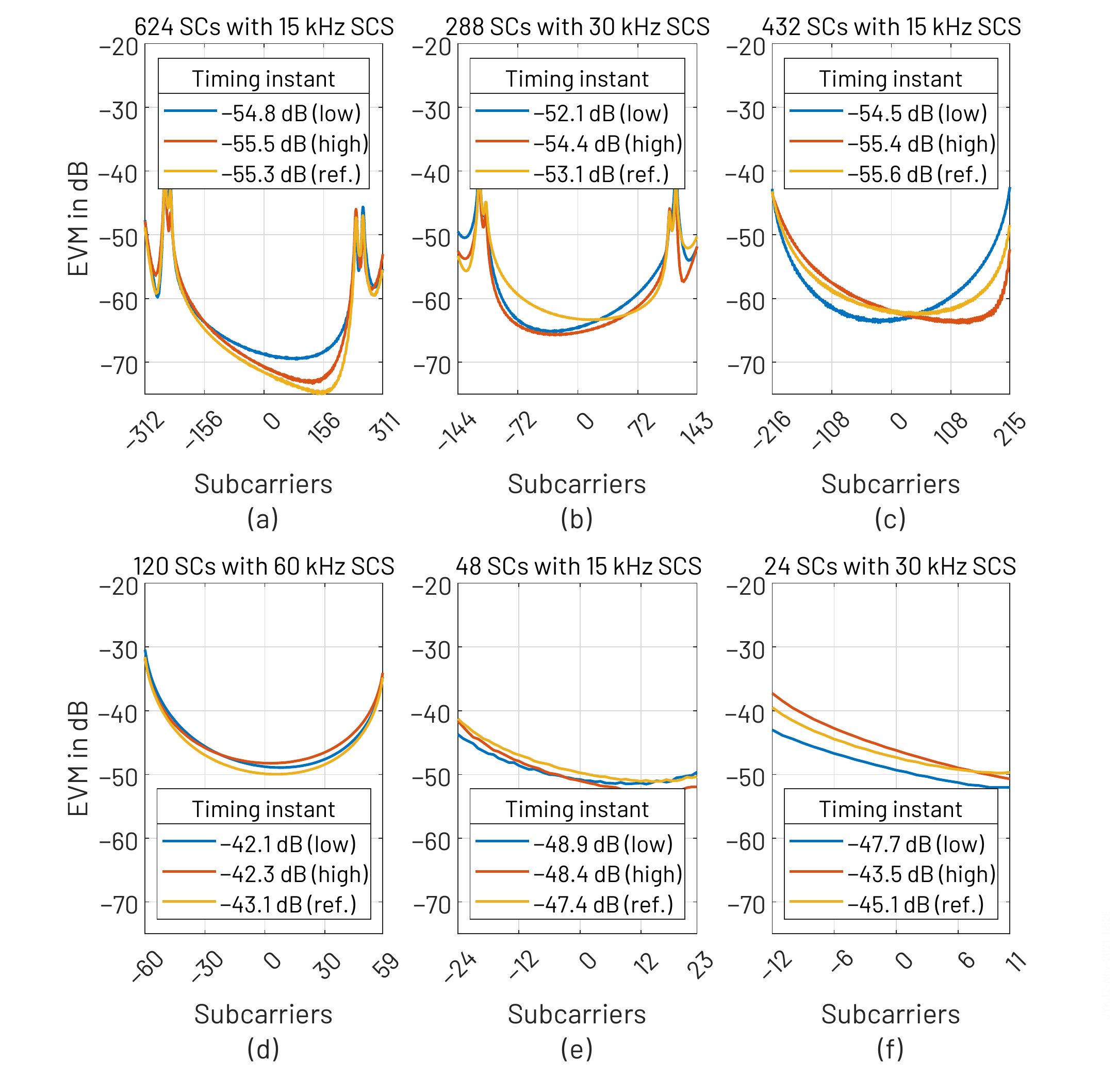}
  \else  
    \includegraphics[
    trim=30 0 45 0,clip,width=\columnwidth]{Figs/EVM_M6_10MHz_new.pdf} 
  \fi    
  \caption{\acs{evm} on each symbol in Example C. (a) 624 \acp{sc} with \SI{15}{kHz} \ac{scs}. (b) 288 \acp{sc} with \SI{30}{kHz} \ac{scs}. (c) 432 \acp{sc} with \SI{15}{kHz} \ac{scs}. (d) 120 \acp{sc} with \SI{60}{kHz} \ac{scs}. (e) 48 \acp{sc} with \SI{15}{kHz} \ac{scs}. (f) 24 \acp{sc} with \SI{30}{kHz} \ac{scs}.}    
  \label{fig:Ex3:EVM}  
\end{figure} 
 
\subsection{Adjustable BWPs}
In this example, we demonstrate the flexibility of the proposed scheme in the case where \ac{fc}-based filtering is reconfigured for each symbol. In this case, we have two variable subbands: First subband has \SI{15}{kHz} \ac{scs} and $L_{\text{act},0,n}=192$ active \acp{sc} for $n=0,1,\dots,6$ while second has \SI{30}{kHz} \ac{scs} and $L_{\text{act},1,n}=72$ active \acp{sc} for $n=0,1,\dots,13$. The center frequency of the first subband is adjusted as $f_{0,n}^\text{(center)} = 64(n-3)\times\SI{15}{kHz}$ for $n=0,1,\dots,6$ and the center frequency of the second subband is  $f_{1,n}^\text{(center)}=218\times\SI{15}{kHz}$ for $n=0,1,\dots,6$ and $f_{1,n}^\text{(center)}=-218\times\SI{15}{kHz}$ for $n=7,8,\dots,13$. This configuration is depicted in Fig.~\ref{fig:Ex4:TFA}.
 
The \ac{psd} of resulting \ac{fc}-based filtered-\ac{ofdm} waveform is shown in Fig.~\ref{fig:Ex4:PSD} while the corresponding \acp{evm} are shown in Fig.~\ref{fig:Ex4:EVM}. As seen from these figures, the performance of the proposed scheme meets the requirements even at this most challenging scenario.
 
\begin{figure}[t!]                 
  \centering   
  \ifonecolumn 
    \includegraphics[width=0.65\columnwidth]{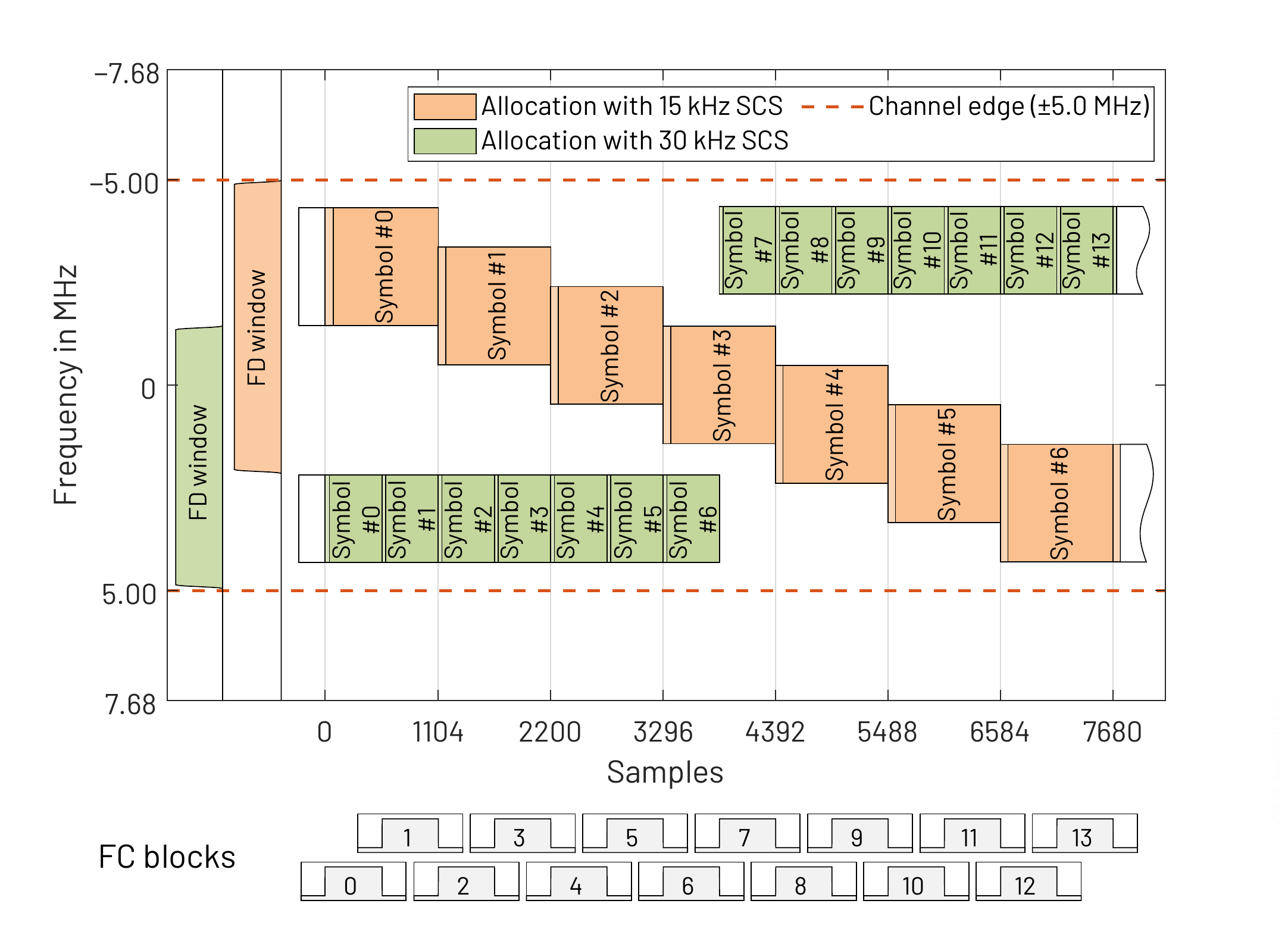}
  \else   
    \includegraphics[
    trim=20 0 50 0,clip,width=0.95\columnwidth]{Figs/TFA_M5_10MHz.pdf} 
  \fi  
  \caption{Time-frequency allocation in Example D with two variable subbands. The frequency-domain windows during the two first \ac{fc} blocks (first symbol with \SI{15}{kHz} \ac{scs} and two first symbols with \SI{30}{kHz} \ac{scs}) are illustrated on the left hand side of the figure.}    
  \label{fig:Ex4:TFA} 
\end{figure}  

\begin{figure}[t!]                 
  \centering   
  \ifonecolumn 
    \includegraphics[width=0.65\columnwidth]{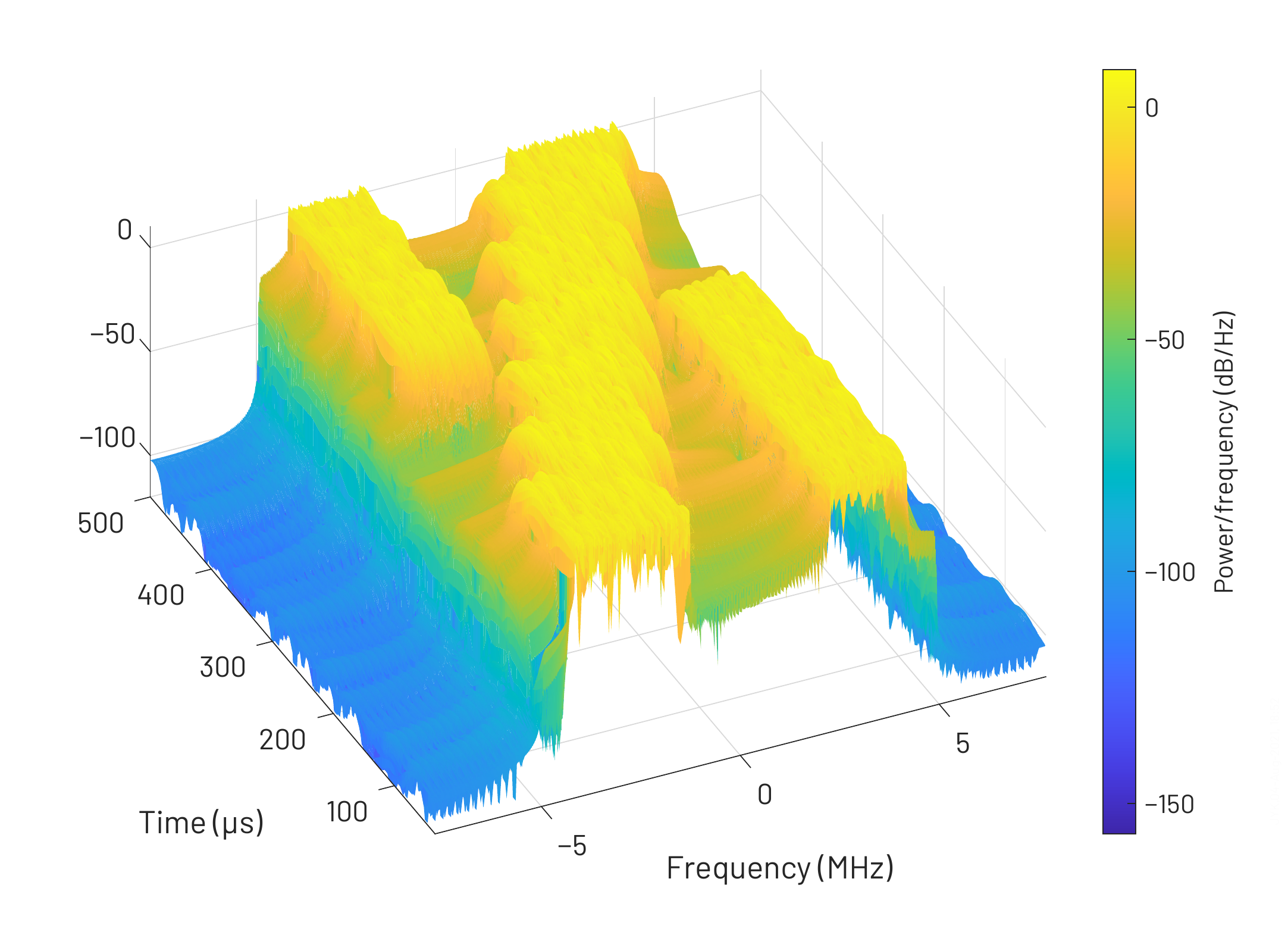}
  \else  
    \includegraphics[
    trim=30 0 0 0,clip,width=\columnwidth]{Figs/PSD10MHzUltim.pdf} 
  \fi  
  \caption{\acs{psd} in Example D with two variable subbands.}    
  \label{fig:Ex4:PSD} 
\end{figure}  
 
\begin{figure}[t!]                 
  \centering   
  \ifonecolumn 
    \includegraphics[width=0.65\columnwidth]{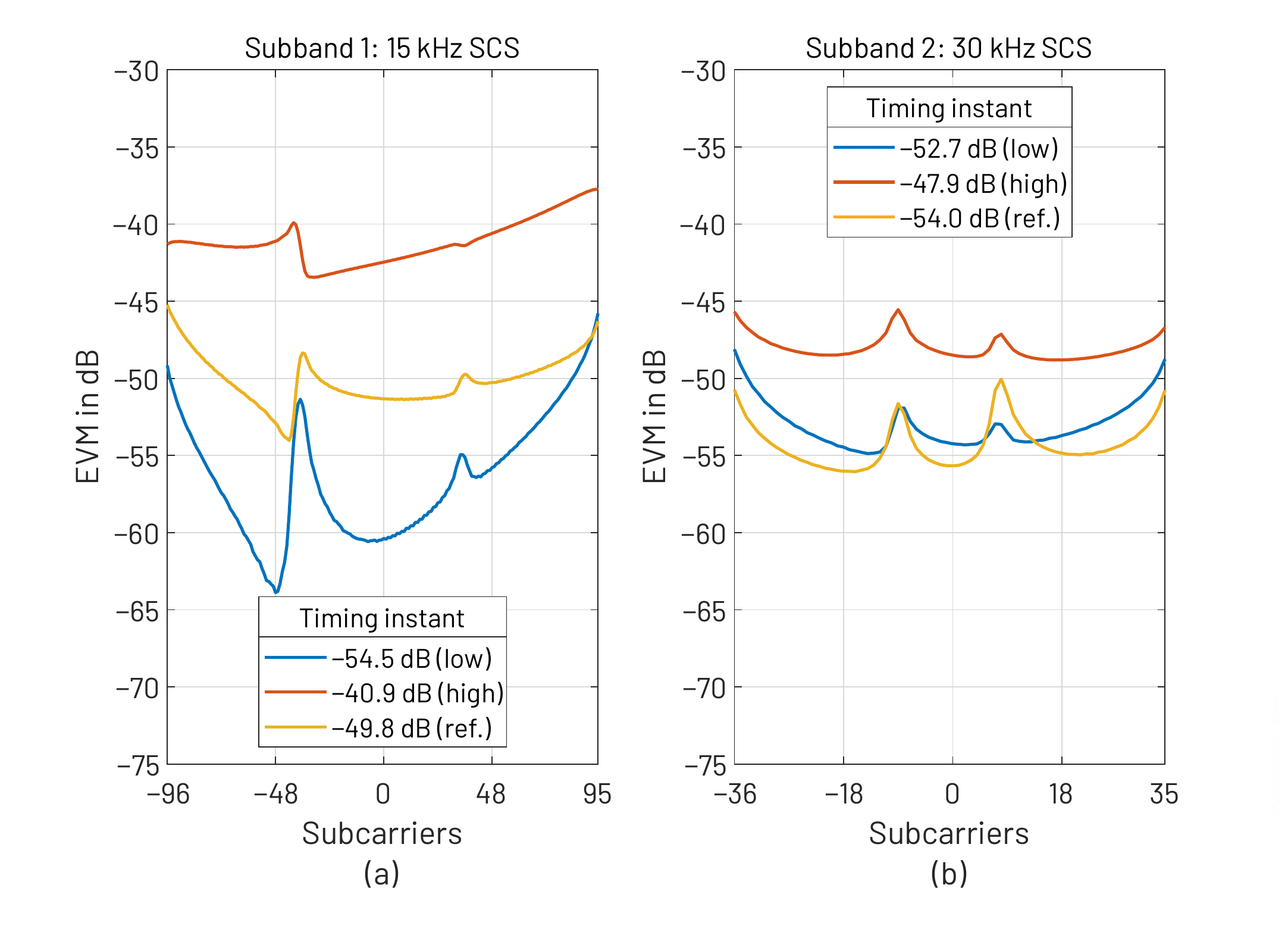}
  \else  
    \includegraphics[
    trim=30 0 55 0,clip,width=\columnwidth]{Figs/EVM_M5_10MHz.pdf} 
  \fi  
  \caption{\acs{evm} on each of two subbands in Example D. (a) \ac{bwp} with \SI{15}{kHz} \ac{scs} and 192 active \acp{sc}. (b) \ac{bwp} with \SI{30}{kHz} \ac{scs} and 72 active \acp{sc}.}    
  \label{fig:Ex4:EVM} 
\end{figure} 

The presented solution allows unseen flexibility in supporting changing allocations in mixed-numerology scenarios with \ac{ofdm} symbol resolution. This allows to support all envisioned use cases for \ac{5gnr} and provides a flexible starting point for \ac{6g} development.

\section{Conclusions}
\label{sec:conclucions}
In this article, continuous symbol-synchronized \acl{fc}-processing scheme was proposed, with particular emphasis on the physical-layer processing in \acl{5gnr} and beyond mobile radio networks. The proposed scheme was shown to offer various benefits over the basic continuous and discontinuous processing models, especially in providing excellent performance in reducing the unwanted emissions and inter-numerology interference in \ac{5g-nr} mixed-numerology scenarios while keeping in-band interference level well below the requirements stated in \ac{3gpp} specifications. Both dynamic and static filtering configurations are supported for all numerologies simultaneously providing greatly improved flexibility over the \ac{fc}-processing schemes proposed earlier. The beneﬁts are particularly important in speciﬁc application scenarios, like transmission of single or multiple narrow subbands, or in mini-slot type transmission, which is a core element in the ultra-reliable low-latency transmission service of \ac{5g-nr} networks. 
 
%
\begin{acronym}[FBMC/OQAM2]
\acro{3gpp}[3GPP]{third generation partnership project}
\acro{4g}[4G]{fourth generation} 
\acro{5g-nr}[5G-NR]{fifth-generation new radio}
\acro{5gnr}[5G NR]{fifth-generation new radio}
\acro{5g}[5G]{fifth generation}
\acro{6g}[6G]{sixth generation}
\acro{F-ofdm}[F-OFDM]{subband filtered CP-OFDM}
\acro{aclr}[ACLR]{adjacent channel leakage power ratio}
\acro{mmwave}[mmWave]{millimeter-wave}
\acro{adsl}[ADSL]{asymmetric digital subscriber line}
\acro{afb}[AFB]{analysis filter bank} 
\acro{af}[AF]{amplify-and-forward} 
\acro{am/am}[AM/AM]{amplitude modulation/amplitude modulation \acroextra{[NL PA models]}}
\acro{am/pm}[AM/PM]{amplitude modulation/phase modulation \acroextra{[NL PA models]}} 
\acro{amr}[AMR]{adaptive multi-rate}
\acro{app}[APP]{a posteriori probability} 
\acro{ap}[AP]{access point} 
\acro{ini}[INI]{inter-numerology interference} 
\acro{awgn}[AWGN]{additive white Gaussian noise}
\acro{b-pmr}[B-PMR]{broadband PMR} 
\acro{bcjr}[BCJR]{Bahl-Cocke-Jelinek-Raviv \acroextra{algorithm}}
\acro{ber}[BER]{bit error rate} 
\acro{bicm}[BICM]{bit-interleaved coded modulation} 
\acro{blast}[BLAST]{Bell Labs layered space time \acroextra{[code]}} 
\acro{bler}[BLER]{block error rate}
\acro{bpsk}[BPSK]{binary phase-shift keying} 
\acro{br}[BR]{bin resolution}
\acro{bs}[BS]{bin spacing}
\acro{BS}[BS]{base station}
\acro{cazac}[CAZAC]{constant amplitude zero auto-correlation}
\acro{cb-fmt}[CB-FMT]{cyclic block-filtered multitone}
\acro{ccc}[CCC]{common control channel}
\acro{ccdf}[CCDF]{complementary cumulative distribution function}
\acro{cdf}[CDF]{cumulative distribution function}
\acro{cdma}[CDMA]{code-division multiple access}
\acro{cfo}[CFO]{carrier frequency offset} 
\acro{cfr}[CFR]{channel frequency response} 
\acro{ch}[CH]{cluster head}
\acro{cir}[CIR]{channel impulse response} 
\acro{cma}[CMA]{constant modulus algorithm} 
\acro{cna}[CNA]{constant norm algorithm}
\acro{cp-ofdm}[CP-OFDM]{cyclic prefix orthogonal frequency-division multiplexing}
\acro{cpu}[CPU]{central processing unit} 
\acro{cp}[CP]{cyclic prefix} 
\acro{cqi}[CQI]{channel quality indicator} 
\acro{crlb}[CRLB]{Cram\'er-Rao lower bound}
\acro{crn}[CRN]{cognitive radio network}
\acro{crs}[CRS]{cell-specific reference signal}
\acro{cr}[CR]{cognitive radio} 
\acro{csir}[CSIR]{channel state information at the receiver}
\acro{csit}[CSIT]{channel state information at the transmitter}
\acro{csi}[CSI]{channel state information}
\acro{d2d}[D2D]{device-to-device} 
\acro{dc}[DC]{direct current}
\acro{dfe}[DFE]{decision feedback equalizer} 
\acro{dft-s-ofdm}[DFT-s-OFDM]{DFT-spread-OFDM}
\acro{dft}[DFT]{discrete Fourier transform} 
\acro{df}[DF]{decode-and-forward} 
\acro{dl}[DL]{downlink} 
\acro{dmo}[DMO]{direct mode operation}
\acro{dmrs}[DMRS]{demodulation reference signals}
\acro{dsa}[DSA]{dynamic spectrum access} 
\acro{dzt}[DZT]{discrete Zak transform} 
\acro{e-utra}[E-UTRA]{evolved UMTS terrestrial radio access}
\acro{ed}[ED]{energy detector}
\acro{egf}[EGF]{extended Gaussian function}
\acro{embb}[eMBB]{enhanced mobile broadband}
\acro{emse}[EMSE]{excess mean square error}
\acro{em}[EM]{expectation maximization} 
\acro{epa}[EPA]{extended pedestrian-A \acroextra{[channel model]}}
\acro{etsi}[ETSI]{European Telecommunications Standards Institute}
\acro{eva}[EVA]{extended vehicular-A \acroextra{[channel model]}}
\acro{evm}[EVM]{error vector magnitude}
\acro{f-ofdm}[f-OFDM]{filtered OFDM}
\acro{fb-sc}[FB-SC]{filterbank single-carrier} 
\acro{fbmc-coqam}[FBMC-COQAM]{filterbank multicarrier with circular offset-QAM}
\acro{fbmc/oqam}[FBMC/OQAM]{filter bank multicarrier with offset-QAM subcarrier modulation} 
\acro{fbmc}[FBMC]{filter bank multicarrier}
\acro{fb}[FB]{filter bank}
\acro{fc-f-ofdm}[FC-F-OFDM]{FC-based filtered-OFDM}
\acro{fc-fb}[FC-FB]{fast-convolution filter bank}
\acro{fc}[FC]{fast-convolution}
\acro{fdma}[FDMA]{frequency-division multiple access}
\acro{fd}[FD]{frequency-domain}
\acro{fec}[FEC]{forward error correction} 
\acro{fft}[FFT]{fast Fourier transform} 
\acro{fir}[FIR]{finite impulse response}
\acro{flo}[FLO]{frequency-limited orthogonal}
\acro{fmt}[FMT]{filtered multitone} 
\acro{fpga}[FPGA]{field programmable gate array} 
\acro{fs-fbmc}[FS-FBMC]{frequency sampled FBMC-OQAM} 
\acro{fr1}[FR1]{frequency range 1}
\acro{fr2}[FR2]{frequency range 2}
\acro{ft}[FT]{Fourier transform}
\acro{re}[RE]{resource element}
\acro{gb}[GB]{guard band}
\acro{gfdm}[GFDM]{generalized frequency-division multiplexing}
\acro{glrt}[GLRT]{generalized likelihood ratio test}
\acro{gmsk}[GMSK]{Gaussian minimum-shift keying}
\acro{hpa}[HPA]{high power ampliﬁer}
\acro{i/q}[I/Q]{in-phase/quadrature \acroextra{[complex data signal components]}} 
\acro{iam}[IAM]{interference approximation method}
\acro{ibe}[IBE]{in-band emission}
\acro{obue}[OBUE]{operating band unwanted emissions}
\acro{ibi}[IBI]{in-band interference}
\acro{ibo}[IBO]{input back-off}
\acro{ici}[ICI]{inter-carrier interference}
\acro{idft}[IDFT]{inverse discrete Fourier transform}
\acro{ifft}[IFFT]{inverse fast Fourier transform}
\acro{iid}[i.i.d.]{independent and identically distributed}
\acro{iota}[IOTA]{isotropic orthogonal transform algorithm}
\acro{iot}[IoT]{internet-of-things}
\acro{isi}[ISI]{inter-symbol interference}
\acro{itu-r}[ITU-R]{International Telecommunication Union Radiocommunication \acroextra{sector}}
\acro{itu}[ITU]{International Telecommunication Union}
\acro{kkt}[KKT]{Karush-K\"uhn-Tucker} 
\acro{kpi}[KPI]{key performance indicator} 
\acro{le}[LE]{linear equalizer}
\acro{llr}[LLR]{log-likelihood ratio} 
\acro{lmmse}[LMMSE]{linear minimum mean squared error} 
\acro{lms}[LMS]{least mean squares}
\acro{lpsv}[LPSV]{linear periodically shift variant} 
\acro{lptv}[LPTV]{linear periodically time-varying} 
\acro{ls}[LS]{least squares} 
\acro{lte-a}[LTE-A]{long-term evolution-advanced}
\acro{lte}[LTE]{long-term evolution}
\acro{lut}[LUT]{look-up table} 
\acro{ssb}[SSB]{synchronization signal block} 
\acro{mac}[MAC]{medium access control}
\acro{mai}[MAI]{multiple access interference}
\acro{map}[MAP]{maximum a posteriori} 
\acro{ma}[MA]{multiple access relay channel} 
\acro{mcm}[MCM]{multicarrier modulation}
\acro{mcs}[MCS]{modulation and coding scheme}
\acro{mc}[MC]{multicarrier}
\acro{mer}[MER]{message error rate} 
\acro{mf}[MF]{matched filter}
\acro{mgf}[MGF]{moment generating function}
\acro{mimo}[MIMO]{multiple-input multiple-output}
\acro{miso}[MISO]{multiple-input single-output}
\acro{mlse}[MLSE]{maximum likelihood sequence estimation} 
\acro{ml}[ML]{maximum likelihood} 
\acro{mma}[MMA]{multi-modulus algorithm} 
\acro{mmse}[MMSE]{minimum mean-squared error}
\acro{mmtc}[mMTC]{massive machine-type communications}
\acro{mos}[MOS]{mean opinion score} 
\acro{mrc}[MRC]{maximum ratio combining} 
\acro{mse}[MSE]{mean-squared error} 
\acro{msk}[MSK]{minimum-shift keying} 
\acro{ms}[MS]{mobile station}
\acro{mtc}[MTC]{machine-type communications}
\acro{mud}[MUD]{multiuser detection} 
\acro{mui}[MUI]{multiuser interference}
\acro{music}[MUSIC]{multiple signal classification}
\acro{mu}[MU]{multi-user} 
\acro{nb-iot}[NB-IoT]{narrow-band internet-of-things}
\acro{nbi}[NBI]{narrowband interference}
\acro{nb}[NB]{narrow-band}
\acro{nc-ofdm}[NC-OFDM]{non-contiguous OFDM}
\acro{nc}[NC]{non-contiguous}
\acro{nl}[NL]{non-linear} 
\acro{nmse}[NMSE]{normalized mean-squared error} 
\acro{npr}[NPR]{near perfect reconstruction}
\acro{nr}[NR]{new radio}
\acro{obo}[OBO]{output back-off} 
\acro{ofdm}[OFDM]{orthogonal frequency-division multiplexing} \acro{ofdma}[OFDMA]{orthogonal frequency-division multiple access} 
\acro{ofdp}[OFDP]{orthogonal finite duration pulse}
\acro{ola}[OLA]{overlap-and-add} 
\acro{ols}[OLS]{overlap-and-save}
\acro{omp}[OMP]{orthogonal matching pursuit}
\acro{oobem}[OOBEM]{out-of-band emission mask}
\acro{oob}[OOB]{out-of-band}
\acro{obw}[OBW]{occupied bandwidth}
\acro{oobe}[OOBE]{out-of-band emission}
\acro{oqam}[OQAM]{offset quadrature amplitude modulation}
\acro{oqpsk}[OQPSK]{offset quadrature phase-shift keying}
\acro{osa}[OSA]{opportunistic spectrum access} 
\acro{pam}[PAM]{pulse amplitude modulation}
\acro{papr}[PAPR]{peak-to-average power ratio}
\acro{pa}[PA]{power amplifier} 
\acro{pci}[PCI]{perfect channel information}
\acro{per}[PER]{packet error rate} 
\acro{pf}[PF]{proportional fair} 
\acro{phy}[PHY]{physical layer}  
\acro{plc}[PLC]{power line communications}
\acro{pmr}[PMR]{professional (or private) mobile radio} 
\acro{pdsch}[PDSCH]{physical downlink shared channel}
\acro{ppdr}[PPDR]{public protection and disaster relief}
\acro{prb}[PRB]{physical resource block}
\acro{prose}[ProSe]{proximity services} 
\acro{pr}[PR]{perfect reconstruction}
\acro{psd}[PSD]{power spectral density} 
\acro{psk}[PSK]{phase-shift keying}
\acro{pswf}[PSWF]{prolate spheroidal wave function}
\acro{pts}[PTS]{partial transmit sequence}
\acro{ptt}[PTT]{push-to-talk} 
\acro{pucch}[PUCCH]{physical uplink control channel}
\acro{pusch}[PUSCH]{physical uplink shared channel}
\acro{pu}[PU]{primary user}
\acro{qam}[QAM]{quadrature amplitude modulation}
\acro{qoe}[QoE]{quality of experience} 
\acro{qos}[QoS]{quality of service}
\acro{qpsk}[QPSK]{quadrature phase-shift keying}
\acro{ram}[RAM]{random access memory} 
\acro{ran}[RAN]{radio access network}
\acro{rat}[RAT]{radio access technology} 
\acro{rbg}[RBG]{resource block group}
\acro{rbw}[RBW]{resolution bandwidth}
\acro{pss}[PSS]{primary synchronization signal}
\acro{sss}[SSS]{secondary synchronization signal}
\acro{pbch}[PBCH]{physical broadcast channel}
\acro{dmrs}[DMRS]{demodulation reference signal}
\acro{bwp}[BWP]{bandwidth part}
\acro{mbw}[MBW]{measurement bandwidth}
\acro{rb}[RB]{resource block}
\acro{rc}[RC]{raised cosine} 
\acro{rf}[RF]{radio frequency} 
\acro{rls}[RLS]{recursive least squares}  
\acro{rms}[RMS]{root mean squared \acroextra{[error]}}
\acro{roc}[ROC]{receiver operating characteristics}
\acro{rrc}[RRC]{square root raised cosine}
\acro{rrm}[RRM]{radio resource management}
\acro{rssi}[RSSI]{received signal strength indicator}
\acro{rx}[RX]{receiver}
\acro{sblr}[SBLR]{subband leakage ratio}
\acro{sc-fde}[SC-FDE]{single-carrier frequency-domain equalization} 
\acro{sc-fdma}[SC-FDMA]{single-carrier frequency-division multiple access} 
\acro{scs}[SCS]{subcarrier spacing}
\acro{sc}[SC]{single-carrier}
\acro{sc}[SC]{subcarriers}
\acro{sdma}[SDMA]{space-division multiple access}
\acro{sdm}[SDM]{space-division multiplexing} 
\acro{sdr}[SDR]{software defined radio}
\acro{sel}[SEL]{soft envelope limiter} 
\acro{ser}[SER]{symbol error rate} 
\acro{sfbc}[SFBC]{space frequency block code}
\acro{sfb}[SFB]{synthesis filter bank}
\acro{sic}[SIC]{successive interference cancellation}
\acro{simo}[SIMO]{single-input multiple-output}
\acro{sinr}[SINR]{signal-to-interference-plus-noise ratio}
\acro{sir}[SIR]{signal-to-interference ratio}
\acro{siso}[SISO]{single-input single-output}
\acro{slm}[SLM]{selected mapping}
\acro{slnr}[SLNR]{signal-to-leakage-plus-noise ratio}
\acro{sm}[SM]{spatial multiplexing}
\acro{sndr}[SNDR]{signal-to-noise-plus-distortion ratio}
\acro{snr}[SNR]{signal-to-noise ratio} 
\acro{softio}[SI-SO]{soft-input soft-output}
\acro{sqp}[SQP]{sequential quadratic programming}
\acro{sspa}[SSPA]{solid-state power amplifiers}
\acro{ss}[SS]{spectrum sensing} 
\acro{sem}[SEM]{spectral emission mask} 
\acro{stbc}[STBC]{space time block code}
\acro{stbicm}[STBICM]{space-time bit-interleaved coded modulation}
\acro{stc}[STC]{space-time coding} 
\acro{sthp}[STHP]{spatial Tomlinson-Harashima precoder} 
\acro{su}[SU]{secondary users}
\acro{svd}[SVD]{singular value decomposition}
\acro{tbw}[TBW]{transition-band width}
\acro{tdd}[TDD]{time-division duplex}
\acro{tdl}[TDL]{tapped-delay line}
\acro{tdma}[TDMA]{time-division multiple access}
\acro{td}[TD]{time-domain}
\acro{teds}[TEDS]{TETRA enhanced data service}
\acro{tetra}[TETRA]{terrestrial trunked radio}
\acro{tfl}[TFL]{time-frequency localization} 
\acro{tgf}[TGF]{tight Gabor frame} 
\acro{tlo}[TLO]{time-limited orthogonal}
\acro{tmo}[TMO]{trunked mode operation} 
\acro{tmux}[TMUX]{transmultiplexer}
\acro{to}[TO]{tone offset}
\acro{tr}[TR]{tone reservation} 
\acro{tsg}[TSG]{technical specification group}
\acro{tx}[TX]{transmitter} 
\acro{ue}[UE]{user equipment} 
\acro{uf-ofdm}[UF-OFDM]{universal filtered OFDM}
\acro{ula}[ULA]{uniform linear array} 
\acro{ul}[UL]{uplink} 
\acro{urllc}[URLLC]{ultra-reliable low-latency communications}
\acro{v-blast}[V-BLAST]{vertical Bell Laboratories layered space-time  \acroextra{[code]}} 
\acro{veh-a}[Veh-A]{vehicular-A \acroextra{[channel model]}} 
\acro{veh-b}[Veh-B]{vehicular-B \acroextra{[channel model]}} 
\acro{wimax}[WiMAX]{worldwide interoperability for microwave access} 
\acro{wlan}[WLAN]{wireless local area network} 
\acro{wlf}[WLF]{widely linear filter}
\acro{wola}[WOLA]{weighted overlap-add} 
\acro{wola}[WOLA]{windowed overlap-and-add} 
\acro{zf}[ZF]{zero-forcing} 
\acro{zfe}[ZFE]{zero-forcing equalizar} 
\acro{zp}[ZP]{zero prefix}  
\acro{zt}[ZT]{Zak transform}
\end{acronym}

 

\def\baselinestretch{1}
\bibliographystyle{IEEEtran} 
\bibliography{References} 
 
\end{document}
 

(defun ats-latex-buffer ()
  (if (eq major-mode 'latex-mode)
      (TeX-run-LaTeX   
       (TeX-master-file) 
       (concat "pdflatex \\\\nonstopmode\\\\input " (TeX-master-file)) 
       (TeX-master-file))))
(add-hook 'after-save-hook 'ats-latex-buffer )